\documentclass[useAMS,usenatbib]{mn2e}

\usepackage{graphicx}
\usepackage{amssymb}
\usepackage{times}
\usepackage{url}
\def\be{\begin{equation}}
\def\ee{\end{equation}}
\def\bea{\begin{eqnarray*}}
\def\eea{\end{eqnarray*}}

\def\url#1{\expandafter\string\csname #1\endcasname}
\def\kms{km s$^{-1}$}

\def\C2{[C~II]}
\def\micron{$\mu$m}
\def\HST{\textit{HST}}
\def\SMA{\textit{SMA}}
\def\Hubble{\textit{Hubble}}
\def\Spitzer{\textit{Spitzer}}

\def\Herschel{\textit{Herschel}}

\def\mnras{MNRAS}
\def\aap{A\&A}
\def\araa{ARA\&A}
\def\aj{AJ}
\def\apjl{ApJ}
\def\apj{ApJ}
\def\apjs{ApJS}

\def\pasj{PASJ}
\def\nat{Nature}

\title[CO and C II Lines in the z=4.3 SMG COSMOS AzTEC-1]{Early Science with the Large Millimeter Telescope: CO and [C~II] Emission in the z=4.3 AzTEC J095942.9+022938 (COSMOS AzTEC-1)}

\author[Min S.~Yun and Other]{Min~S.~Yun$^1$\thanks{E-mail: myun@astro.umass.edu},
I.~Aretxaga$^2$,
M.~A.~Gurwell$^3$ ,
D.~H.~Hughes$^2$,
A.~Monta{\~n}a$^{4,2}$ ,
G.~Narayanan$^1$,
\newauthor
D.~Rosa~Gonz{\'a}lez$^2$,
D.~S\'{a}nchez-Arg\"{u}elles$^2$,
F.~P.~Schloerb$^1$,
R.~L.~Snell$^1$,
O. Vega$^2$, 
\newauthor
G.~W.~Wilson$^1$,
M.~Zeballos$^2$, 
M. Chavez$^2$, 
J.~R.~Cybulski$^1$,
T.~D\'{i}az-Santos$^5$,
V.~De~la~Luz$^{2,6}$,
\newauthor
N.~Erickson$^1$,
D.~Ferrusca$^2$ ,
H.~B.~Gim$^1$,
M.~H.~Heyer$^1$,
D.~Iono$^7$,
A.~Pope$^1$,
S.~M.~Rogstad$^1$,
\newauthor
K. S. Scott$^8$, 
K.~Souccar$^1$,
E. Terlevich$^2$, 
R. Terlevich$^2$, 
D. Wilner$^3$,
J. ~A.~Zavala$^2$ \\
$^{1}$Department of Astronomy, University of Massachusetts, Amherst, MA 01003, USA\\
$^2$Instituto Nacional de Astrof\'{i}sica, \'{O}ptica y Electr\'{o}nica, Tonantzintla, Luis Enrique Erro 1, Sta. Ma. Tonantzintla, Puebla, M\'{e}xico\\
$^{3}$Harvard-Smithsonian Center for Astrophysics, 60 Garden Street, Cambridge, MA 02138, USA\\
$^{4}$Consejo Nacional de Ciencia y Tecnolog\'{i}a, Av. Insurgentes Sur 1582, Col. Cr\'{e}dito Constructor, Del. Benito Ju\'{a}rez, C.P.: 03940, M\'{e}xico, D.F.\\
$^{5}$Nucleo de Astronomia de la Facultad de Ingenieria, Universidad Diego Portales, Av. Ejercito Libertador 441, Santiago, Chile\\
$^{6}$SCiESMEX, Instituto de Geof\'{i}sica, Unidad Michoacan, Universidad Nacional Aut\'{o}noma de Mexico, Morelia, Michoacan, Mexico. CP 58190.\\
$^{7}$National Astronomical Observatory of Japan, 2-21-1 Osawa, Mitaka,Tokyo 181-8588, Japan\\
$^{8}$National Radio Astronomy Observatory, 520 Edgemont Road, Charlottesville, VA, USA\\
}

\begin{document}

\date{\today}

\pagerange{\pageref{firstpage}--\pageref{lastpage}} \pubyear{2015}

\maketitle

\label{firstpage}

\begin{abstract}

Measuring redshifted CO line emission is an unambiguous method for obtaining an accurate redshift and total cold gas content of optically faint, dusty starburst systems.  Here, we report the first successful spectroscopic redshift determination of  AzTEC J095942.9+022938 (``COSMOS AzTEC-1"), the brightest 1.1mm continuum source found in the AzTEC/JCMT survey \citep{scott08}, through a clear detection of the redshifted CO (4-3) and CO (5-4) lines using the Redshift Search Receiver on the Large Millimeter Telescope.  The CO redshift of $z=4.3420\pm0.0004$ is confirmed by the detection of the redshifted 158 \micron\  \C2\ line using the Submillimeter Array.  The new redshift and \Herschel\ photometry yield $L_{FIR}=(1.1\pm0.1)\times 10^{13} L_\odot$ and $SFR\approx 1300\, M_\odot$ yr$^{-1}$.  Its molecular gas mass derived using the ULIRG conversion factor is $1.4\pm0.2 \times 10^{11} M_\odot$ while the total ISM mass derived from the 1.1mm dust continuum is $3.7\pm0.7 \times 10^{11} M_\odot$ assuming $T_d=35$ K.  
Our dynamical mass analysis suggests that the compact gas disk ($r\approx 1.1$ kpc, inferred from dust continuum and SED analysis) has to be nearly face-on, providing a natural explanation for the uncommonly bright, compact stellar light seen by the \HST.  
The \C2\ line luminosity $L_{[C\, II]}= 7.8\pm1.1  \times 10^9 L_\odot$ is remarkably high, but it is only 0.04 per cent of the total IR luminosity.  AzTEC COSMOS-1 and other high redshift sources with a spatially resolved size extend the tight trend seen between \C2/FIR ratio and $\Sigma_{FIR}$ among IR-bright galaxies reported by \citet{diaz13} by more than an order of magnitude, supporting the explanation that the higher intensity of the IR radiation field is responsible for the ``\C2\ deficiency" seen among luminous starburst galaxies.

\end{abstract}

\begin{keywords}
galaxies: high-redshift -- galaxies: starburst -- galaxies: distances and redshifts --  galaxies: individual: AzTEC J095942.9+022938 -- submillimetre: galaxies -- radio lines: ISM 
\end{keywords}

\section{Introduction}

Recent studies of cosmic star formation history and galaxy mass build-up have shown a remarkably tight correlation between star formation rate (SFR) and stellar mass ($M_*$), also known as star formation ``main sequence", for galaxies with $M_*$ up to $10^{11} M_\odot$ extending out to $z\sim 6$ \citep[see][and references therein]{steinhardt14,salmon15}.  A substantial population of quiescent galaxies with $M_*\ge 10^{10-11} M_\odot$ are also found to $z\sim4$, suggesting rapid formation and quenching of massive galaxies at $z\sim 6$ or earlier \citep{whitaker13,straatman14}.  Given the constraints on rapid formation and cessation of stellar mass build-up and their compact morphology, intense starbursts and feedback driven by a rapid gas accretion are thought to be important in this process \citep[see][and references therein]{williams14}. 

The submillimetre galaxies (SMGs) are natural laboratories for testing this hypothesis and probing the details of the physical processes that govern this rapid build-up and quenching of massive galaxies.  SMGs are identified by their large FIR luminosity, which is widely interpreted to be powered by intense star formation with $SFR\ge 10^{2-3} M_\odot$ yr$^{-2}$ \citep{blain02,yun12,kirkpatrick12}.  Wide area surveys by the \Spitzer\ and \Herschel\ Space Telescopes have shown that these luminous IR galaxies account for a significant fraction of the Cosmic IR background \citep{penner11,bethermin12}, suggesting that they are an important component of the cosmic mass build-up history at $z\ge1$ \citep{lefloch05,caputi07,magnelli11}.  Because of their faintness in the optical bands, their precise redshift distribution is poorly determined (see below), but \citet{toft14} have found that the well established population of massive ($M_*>10^{11}M_\odot$) compact quiescent galaxies at $z\sim2$ can be fully accounted by the known SMGs at $3<z<6$ in terms of their abundance and stellar population, and \citet{simpson14} reproduce the local elliptical luminosity function by passively evolving the population of bright SMGs.  

The study of the brightest SMG found in the COSMOS \citep{scoville07} field, AzTEC J095942.9+022938 (``COSMOS AzTEC-1" hereafter), exemplifies the major challenge behind making this important connection between SMGs and the rapid build-up of stellar mass in galaxies.  First discovered by the AzTEC COSMOS survey using the James Clerk Maxwell Telescope \citep{scott08}, COSMOS AzTEC-1 is one of the brightest SMGs known and is particularly well studied because of the extensive deep multi-wavelength data readily available in the COSMOS field \citep[see][]{smolcic11}.  Unlike many other SMGs (including those discovered by \Herschel) that suffer from low angular resolution of single dish telescopes and source blending, the location of this SMG is known to better than $0.1\arcsec$ accuracy because of a dedicated interferometric imaging survey done using the Submillimeter Array (SMA) by \citet{younger07}.  COSMOS AzTEC-1 is the only object among the 7 AzTEC sources imaged with SMA by Younger et al. that has an unambiguous optical counterpart, and this relatively bright ($m=25.3$ mag [AB] in the \HST\ F814W band) source is extremely compact, $\sim$0.2\arcsec\ ($\sim$1.5 kpc) in diameter, comparable to the sizes of the massive compact galaxies found at $z=2\sim 4$.  A 4 hour long exposure with DEIMOS on Keck II telescope by \citet{smolcic11} did not yield any emission lines, and the continuum break near 6700 \AA\ was interpreted as the blue cutoff of Ly-$\alpha$ at $z=4.65$.  Also, using 31 NUV-NIR photometric measurements, Smol{\v c}i{\'c} et al. derived a photometric redshift of $z=4.64$ with a secondary peak at $z=4.44$.  Their attempt to confirm this redshift by CO spectroscopy using CARMA and PdBI interferometers failed to detect any CO emission in the redshift range of $4.56 <z< 4.76$ and $4.94<z<5.02$.  Later, \citet{iono12} expanded the CO line search using the Nobeyama 45-m Telescope to $4.38<z<4.56$, but they also failed to detect a CO line.  Therefore, the redshift of this arguably the best studied AzTEC SMG in the COSMOS field still lacks a spectroscopic confirmation despite nearly 10 years of efforts using some of the most powerful astronomical facilities available.  With the exception of the SMG COSMOS AzTEC-3, which is recently shown to be part of a large scale structure at $z=5.3$ \citep{riechers10,riechers14}, the situation is essentially the same for the remaining AzTEC sources imaged by the SMA as well -- all are expected to be at $z\gg3$ because of their faintness in the optical and the radio bands, with a much worse prospect of yielding a spectroscopic redshift.

In this paper, we report the first successful spectroscopic redshift determination of COSMOS AzTEC-1 obtained using the Redshift Search Receiver (RSR) on the Large Millimeter Telescope  Alfonso Serrano (LMT), which is a ultra-wide bandwidth spectrometer designed to conduct a blind search for redshifted CO lines from molecular gas rich galaxies.  We also report the confirmation of the CO redshift through the detection of the redshifted 158 \micron\ \C2\ line using the Submillimeter Array.   We interpret the observed CO and \C2\ luminosity in terms of a highly concentrated and intense starburst, fueled by the CO and \C2\ emitting gas and its properties, including the  ``\C2\ deficiency" in COSMOS AzTEC-1.  Throughout this paper, we assume flat $\Lambda$CDM cosmology with $\Omega_M=0.3$ and $H_0=70$ km s$^{-1}$ Mpc$^{-1}$ and Kroupa initial mass function \citep[IMF;][]{kroupa01}.\footnote{Stellar mass in Kroupa IMF is 38 per cent smaller than that of the Salpeter IMF -- i.e., $M_*(Kroupa) = 0.62 M_*(Salpeter)$.}.

\section{Observations}

\subsection{Redshift Search Receiver Observations on the LMT}

The Redshift Search Receiver observations of COSMOS AzTEC-1 were conducted in January and February 2014 as part of the Early Science program at the Large Millimeter Telescope \citep{hughes10}.  The Redshift Search Receiver (RSR) consists of two dual polarization front end receivers that are chopped between the ON and OFF source positions separated by 76$\arcsec$ in Azimuth at 1 kHz rate using a ferrite switch, producing a flat baseline over the entire 38 GHz (73-111 GHz) bandwidth and always integrating on-source \citep[see][for further descriptions of the instrument]{erickson07,chung09}. The ultra-wideband backend spectrometer covers the entire frequency range between 73 and 111 GHz simultaneously with 31 MHz ($R=3000$ or 100 km s$^{-1}$ at 93 GHz) spectral resolution.  During this Early Science phase operation, only the inner 32 metre diameter section of the telescope surface is illuminated, leading to an effective beam size of $20\arcsec$ at 110 GHz and $28\arcsec$ at 75 GHz.
A total of 290 minutes of on-source integrations were obtained over 3 different nights, mostly in excellent weather with $T_{sys}\approx 90$ K ($\tau_{225GHz}=0.05-0.1$).  Telescope pointing was checked every 60-90 minutes by observing the nearby QSO J0909+013.  

Data were reduced and calibrated using DREAMPY (Data REduction and Analysis Methods in PYthon), which is the RSR data reduction pipeline software written by G. Narayanan.  After flagging any data adversely affected by a hardware or software problem, a linear baseline is removed from each spectrum.  The final spectrum shown in Figure~\ref{fig:CO} was obtained by averaging all spectra using the $1/\sigma^2$ weight, and the resulting final rms noise is $\sigma=0.13$ mK.   The measured gain as a function of elevation and frequency using Uranus and MWC349A is flat, $7$ Jy/K (in $T_A^*$ unit) between the elevation range of 30-75 degree, where all observations were made.  

\subsection{Submillimeter Array}

Spectroscopic imaging observations of COSMOS AzTEC-1 were obtained on 1 March 2014 using the using the Submillimeter Array \citep[SMA;][]{ho04}, the 8-element interferometer located near the summit of Mauna Kea, Hawaii.  The SMA was in a close pack configuration with projected baselines ranging from 6 to 45 metre (mean $\sim$21 metre).  The array was operated using two orthogonally polarized SIS receivers on each antenna, each tuned to 355.7 GHz within the 2 GHz wide upper sideband, the expected frequency for the redshifted \C2\ line based on the results of the LMT observations.  The lower sideband was also captured allowing a sensitive measure of the thermal continuum near 345.4 GHz (1850 GHz = 162 \micron\ in the source frame).  The raw spectral resolution was 3.25 MHz uniform over both sidebands, or about 2.75 km s$^{-1}$ around the \C2\ line.   

The spectral response was calibrated using observations of the bright QSO 3C~84, and the flux density scale was calibrated from measurements of the Jovian moon Callisto, known to within ~5 per cent in the submillimetre bands (based upon SMA observations, see ALMA Memo 594\footnote{\url{http://library.nrao.edu/public/memos/alma/memo594.pdf}}).  Observations of the target were interleaved with measurements of QSOs J0909+013 (0.326 Jy) and J1058+015 (2.31 Jy) for use in calibrating the complex gains due to instrumental and atmospheric effects.  The observations were obtained in very good weather, with $\tau_{225GHz} \sim 0.075$.  The total on-source integration time was 5.2 hours.  The complex visibility data were calibrated within the MIR reduction package\footnote{\url{https://www.cfa.harvard.edu/~cqi/mircook.html}}, and the calibrated visibilities were then exported to MIRIAD \citep{sault95}  for resampling to a common spectral grid.

The continuum and the \C2\ spectral line image cube were produced using the Astronomical Image Processing System (AIPS)\footnote{\url{http://www.aips.nrao.edu/index.shtml}} task {\em IMAGR}.   Since no spatial details are expected to be revealed at the angular resolutions achieved (see below), natural weighting was used in the mapping in order to maximize sensitivity.  The 345 GHz continuum image with an effective bandwidth of 2 GHz has a synthesized beam of $5.8\arcsec \times 3.4\arcsec$  ($PA=60^\circ$) and $1\sigma$ noise of 1.4 mJy beam$^{-1}$.  The spectral resolution of the line data is 11.865 MHz (10 km s$^{-1}$ at 355 GHz).  The final \C2\ spectral line cube was produced by first subtracting the continuum from the {\em uv-data} and then averaging over 8 channels with an increment of 4 channels, covering the frequency range between  354.4347 GHz and 356.2381 GHz.  The resulting cube has a spectral resolution of 94.92 MHz with a synthesized beam of $5.7\arcsec \times 3.3\arcsec$  ($PA=59^\circ$) and $1\sigma$ noise of 4.2 mJy beam$^{-1}$ in each channel.  
 
\section{Results}

\subsection{CO Redshift and Line Luminosity \label{sec:CO}}

\begin{table*}
 \caption{Summary of the CO and \C2\ Line Measurements}
 \label{tab:CO}
 \begin{tabular}{@{}lccccccc@{}}
  \hline
%  ID & $\nu_{CO}$ (GHz) & Line & $z_{CO}$ & $\Delta V$ (km/s) & $S\Delta V$ (Jy km/s) & $L'_{CO}$ ($10^{10}$ K km/s pc$^2$) & $M_{H2}$ ($10^{10} M_\odot$) & $M_*$ ($M_\odot$) \\
   Line & $\nu_{CO/[C\, II]}$  & $z_{CO/[C\, II]}$ & $\Delta V$  & $S\Delta V$ & $L_{CO/[C\, II]}$ & $L'_{CO/[C\, II]}$ & $M_{H2}^b$ \\
   & (GHz) & & (km s$^{-1}$) & (Jy km s$^{-1}$) & ($10^8 L_\odot$) & ($10^{10}$ K km s$^{-1}$ pc$^2$) & ($10^{10} M_\odot$)  \\
 \hline
 CO (4--3) & 86.3085  & $4.3418\pm0.0006$ & 380 & $1.75^a\pm0.24$ & $2.5\pm 0.3$ & $7.8\pm 1.1$ & $14\pm 2$ \\       
 CO (5--4)  & 107.8739  & $4.3421\pm0.0006$ & 364 & $1.55^a\pm0.22$ & $2.8\pm 0.4$ & $4.4\pm 0.6$ & $8.8\pm 1.2$ \\       
 \C2\  & 355.8038  & $4.3415\pm0.0003$ & 366 & $13.05\pm0.70$ & $78\pm 5$ & $3.6\pm 0.2$ & -- \\       
 \hline
 \end{tabular}
 
 \medskip
(a) A conversion of 7 Jy/K is adopted to convert the measured antenna temperature in $T_A*$ to flux density, using the calibration factor derived between December 2013 and January 2014.\\
(b) $L'_{CO(1-0)}$ is estimated using the average line ratios for SMGs \citep{carilli13}, and $M_{H2}$ is derived using the ``ULIRG" conversion factor $\alpha_{CO}=0.8 M_\odot$ (K km s$^{-1}$ pc$^2$)$^{-1}$ -- see \S~\ref{sec:mass}.
\end{table*}

% enter f1, f2, fc, amp, fwhm: 85.3 87.3 86.3 .46 .04
% ndata used:           64
% peak =       0.44860582
% center =        86.308478
% sigma =      0.046473783
% offset =     0.0033276898
% FWHM(km/s) =        380.12756
% line integral (K km/s) =       0.24957065     0.033851665       7.3724779
%z(1-0) =       0.33557217   0.00013782048
%z(2-1) =        1.6710933   0.00027563568
%z(3-2) =        3.0065124   0.00041344036
%z(4-3) =        4.3417785   0.00055122925
%z(5-4) =        5.6768405   0.00068899709
%z(6-5) =        7.0116472   0.00082673858

% enter f1, f2, fc, amp, fwhm: 107.4 108.4 107.9 .54 .02
% ndata used:           32
% peak =       0.55726687
% center =        107.87392
% sigma =      0.055641606
% offset =   -0.00041177195
% FWHM(km/s) =        364.13128
% line integral (K km/s) =       0.22134055     0.031274213       7.0774140
%z(1-0) =      0.068573422   0.00011003181
%z(2-1) =        1.1371060   0.00022005942
%z(3-2) =        2.2055570   0.00033007862
%z(4-3) =        3.2738855   0.00044008521
%z(5-4) =        4.3420507   0.00055007500
%z(6-5) =        5.4100117   0.00066004375

\begin{figure}
\includegraphics[width=0.99\columnwidth]{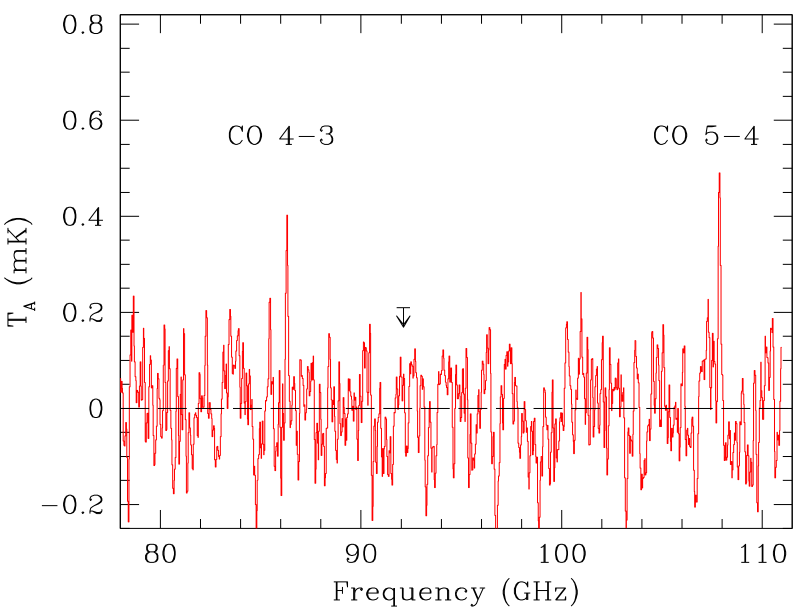}
\caption{RSR spectrum of COSMOS AzTEC-1. The two spectral features well above the noise level are interpreted as CO (4--3) and (5--4) lines at $z=4.342$.  The redshifted 492 GHz [C~I] line should appear at 92.13 GHz (marked with an arrow). }
\label{fig:CO}
\end{figure}

\begin{figure}
\includegraphics[width=0.99\columnwidth]{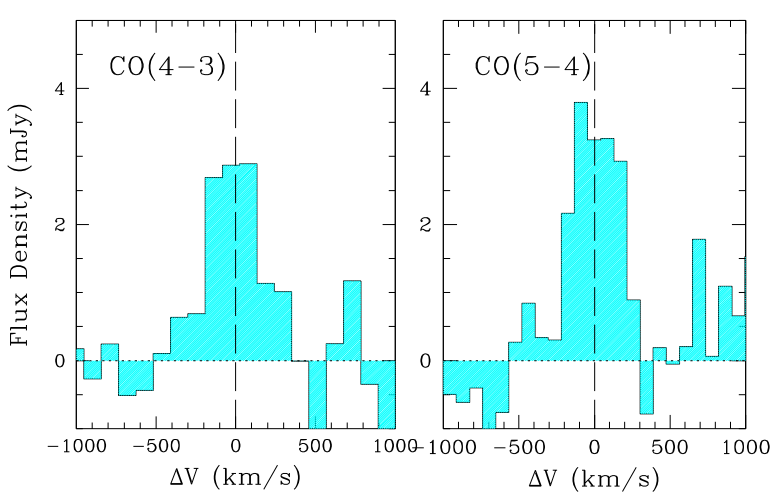}
\caption{Zoom-in views of the two CO lines in the RSR spectrum of COSMOS AzTEC-1.  The x-axis is in velocity offset with respect to the systematic redshift of $z=4.3420$ (vertical long-dashed lines). }
\label{fig:COzoom}
\end{figure}

The final RSR spectrum of COSMOS AzTEC-1 shown in Figure~\ref{fig:CO} has two emission lines clearly above the noise level.  A straightforward interpretation of the spectrum is that these are two redshifted, adjacent rotational transitions of CO, and the separation between the two lines, $\Delta\nu = 21.565$ GHz, corresponds to the expected frequency offset between CO $J=4 \rightarrow 3$ and $J=5\rightarrow 4$ transitions at $z= 4.342$ (see Appendix for a detailed discussion on the redshift determination from an RSR spectrum).   Both lines are fully resolved by the RSR (see Figure~\ref{fig:COzoom}), and the best fit Gaussian parameters are summarized in Table~\ref{tab:CO}.  The CO (4--3) line is centered at $\nu=86.3085$ GHz, corresponding to a redshift of $4.3418\pm0.0006$ while the CO (5--4) line is centered at $\nu=107.8739$ GHz, at a redshift of $4.3421\pm0.0006$.  The best-fit linewidths (FWHM) are 380 \kms\ and 364 \kms, respectively, in good agreement as expected if they are two CO transitions from the same galaxy.   The redshifted 492 GHz [C~I] line, which should appear at 92.130 GHz, is undetected with $S_{[C~I]}/S_{CO(4-3)} \le 0.45$, in line with the measured [C~I] line strengths in other high redshift galaxies \citep[][$S_{[C~I]}/S_{CO(3-2)} \sim 0.3$]{walter11}.

For an ultra-wide spectrum produced by the RSR, the redshift information is also present in weak lines such as [C~I] that are not formally detected individually as well as in bright lines such as CO.  We have developed a method to exploit all spectral information present in the RSR data by cross-correlating the observed spectrum with a theoretical or an empirical spectral template \citep[see][]{yun07}.  A detailed analysis of the cross-correlation amplitude along with the expected CO line multiplicity and the redshift constraints from the radio-millimetric photometric redshift analysis uniquely identifies the $z=4.342$ solution with a total $S/N=9.0$ (see Appendix).  As shown in the zoom-in inset of Figure~\ref{fig:COLadder}, this redshift peak is well-resolved by the cross-correlation analysis with a spread in redshift between 4.338 and 4.347 (FWHM). It is well centered on the redshift of AzTEC-1 derived from fitting the individual CO lines, $z=4.3420\pm0.0004$ (see Table~\ref{tab:CO}), but the width of the distribution is nearly 10 times larger than the uncertainty from the individual line fitting, indicating that the width of the cross-correlation amplitude arises from the finite width of the CO lines ($\sim$375 km s$^{-1}$) rather than reflecting the uncertainty in the redshift determination.  

The CO line redshift of $z=4.3420\pm0.0004$ we derive from the RSR spectrum is significantly lower than the Lyman-$\alpha$ break based redshift of $4.650\pm0.005$ or the optical/IR photometric redshift of $z=4.64^{+0.06}_{-0.08}$ reported by \citet{smolcic11}, and naturally explains why their CARMA and PdBI CO line searches failed.  Their photometric redshift analysis produced a secondary solution at $z=4.44$, which is much closer to our CO redshift.  Although \citet{iono12} had the right idea to search for a CO line near this secondary redshift peak, they missed detecting the CO (5--4) line just outside their search range.   In either case, the redshift adopted by \citet{smolcic11} is close enough to the actual CO redshift that their analysis of stellar mass and IR spectral energy distribution is still mostly valid, and their conclusion that AzTEC-1 is an extremely young ($\le 50$ Myr), massive ($M_* \sim 10^{11} M_\odot$), and compact ($\le2$ kpc) galaxy with a star formation rate of $SFR\sim10^3 M_\odot$ yr$^{-1}$ still holds.

The CO line luminosity $L_{CO}$ in $L_\odot$ can be computed from the measured line integrals and the CO redshift using Eq.~(1) by \citet{solomon97} as,
\begin{equation}
   L_{CO}=1.04\times 10^{-3}S_{CO}\Delta V \nu_0 (1+z)^{-1}D_L^2 ~~[L_\odot] 
\end{equation}
 where
 $S_{CO} \Delta V$ is the measured CO line integral in Jy km s$^{-1}$,
 $\nu_0$            is the rest frequency of the CO transition in GHz, and
 $D_L$ is the luminosity distance in Mpc.  As summarized in Table~\ref{tab:CO},  $L_{CO}$ is $(2.5\pm0.3)\times10^8 L_\odot$ and $(2.8\pm0.4)\times10^8 L_\odot$ for the CO (4--3) and CO (5--4) transitions, respectively.

Total molecular gas mass is related to the quantity $L'_{CO}$ \citep[see Eq.~(3) by][]{solomon97},
\begin{equation}
   L'_{CO}=3.25\times 10^7 S_{CO}\Delta V \nu_{obs}^{-2} (1+z)^{-3}D_L^2 ~~[K\, km\, s^{-1} pc^2]
\end{equation}
 where
  $\nu_{obs} = \nu_0/(1+z)$ is the observed line frequency in GHz.  The derived CO luminosities are $L'_{CO(4-3)}=(7.8\pm1.1)\times 10^{10}$ K km s$^{-1} $ pc$^2$ and $L'_{CO(5-4)}=(4.4\pm0.6)\times 10^{10}$ K km s$^{-1}$ pc$^2$.  These quantities can be converted to molecular gas mass $M_{H2}$ with several assumptions.  Given the large uncertainties involved in this conversion, the total gas mass estimation from $L'_{CO}$ is deferred to a discussion later (see \S~\ref{sec:mass} below).

\subsection{\C2\ Line and 345 GHz Continuum \label{sec:C2}}

\begin{figure}
\includegraphics[width=0.99\columnwidth]{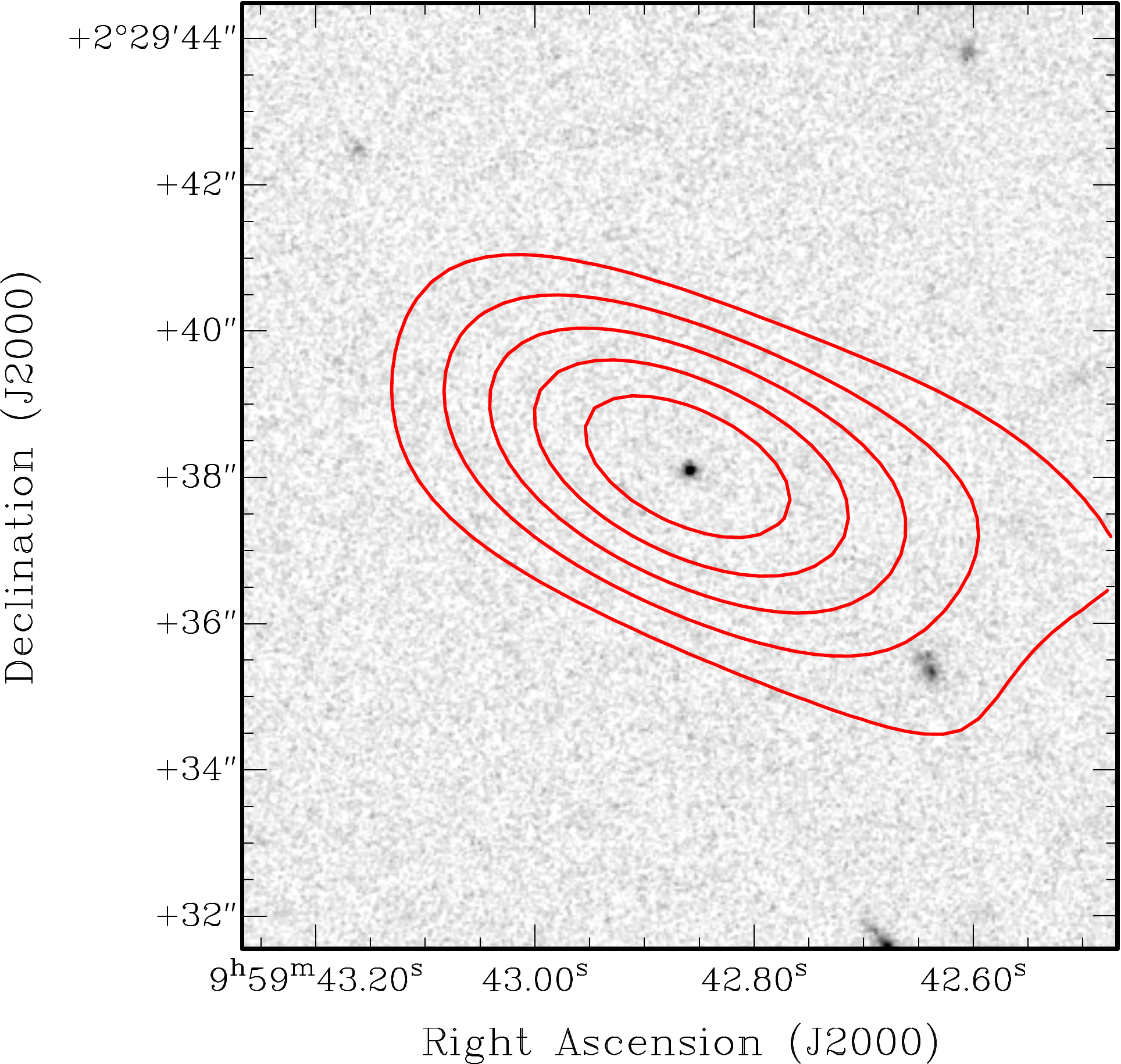}
\caption{SMA 345 GHz continuum image of COSMOS AzTEC-1 at $5.8\arcsec \times 3.4\arcsec$ resolution.  Contours correspond to  2$\sigma$, 4$\sigma$, 6$\sigma$, 8$\sigma$, \& 10$\sigma$ ($\sigma=1.4$ mJy/beam).  The \HST\ $i$-band image is shown in greyscale. }
\label{fig:345continuum}
\end{figure}

\begin{figure}
\includegraphics[width=0.99\columnwidth]{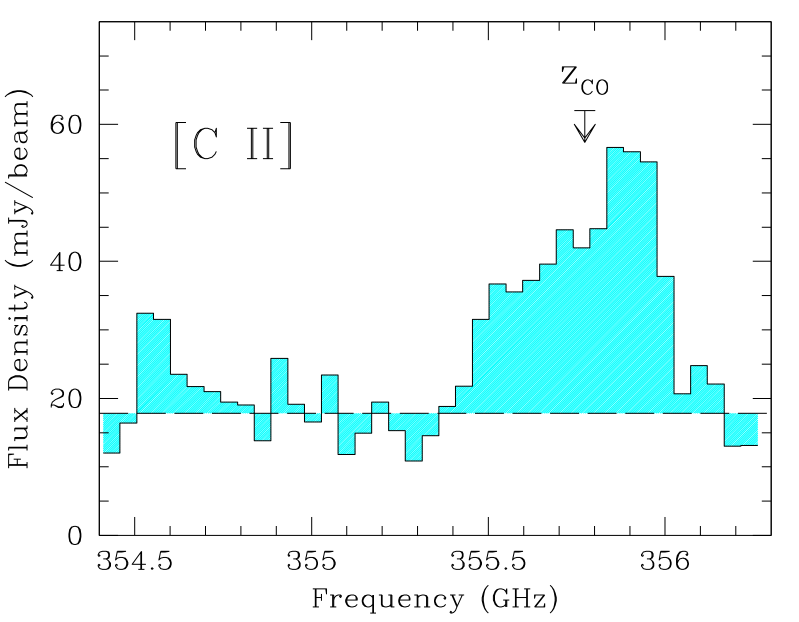}
\caption{A \C2\ spectrum of COSMOS AzTEC-1 obtained using the SMA.  The \C2\ line redshift of $z_{[C\, II]}=4.3415 \pm 0.0003$ and $\Delta V = 366$ \kms\ are in excellent agreement with the RSR CO line measurements (see \S~\ref{sec:CO}).   }
\label{fig:C2}
\end{figure}

While the two CO lines detected with the RSR on the LMT look quite good, confirming the CO redshift of $z=4.342$ and ruling out the possibility of a chance superposition of two unrelated low redshift ($z<3$) CO sources along the same line of sight requires a confirmation with another emission line.  The redshifted \C2\ line ($\nu_0 = 1900.537$ GHz) falls in the middle of the SMA 350 GHz receiver band, which also overlaps with the frequency coverage of the SMA 400 GHz receiver.   The SMA spectrum obtained (Figure~\ref{fig:C2}) shows a clearly detected and fully resolved  \C2\ line near 355.7 GHz with $S/N\sim 15$, confirming the CO redshift.  Three other redshifted \C2\ line sources (the $z=4.7$ QSO BR1202$-$0725 \citep{iono06}, the $z=5.24$ lensed \Herschel\ source HLSJ091828+514223 \citep{rawle14}, and the $z=4.68$ lensed SMG HLS1-MACSJ2043 \citep{zavala15}) have been detected by the SMA before, and this new \C2\ detection of COSMOS AzTEC-1 nicely demonstrates the intrinsic brightness of the \C2\ line and the excellent sensitivity of the SMA for studying high redshift \C2\ sources.

Because the \C2\ line is detected with a significantly higher $S/N$ and a higher spectral resolution than the RSR CO observations, this SMA \C2\ spectrum can reveal much more than simply confirming the redshift of COSMOS AzTEC-1.  The \C2\ line shown in Figure~\ref{fig:C2} is slightly asymmetric, spanning 520 MHz (438 \kms).  The center of the line at full-width-zero-intensity (FWZI) corresponds to $\nu=355.74$ GHz or $z=4.3425$, which agrees very well with the CO redshift.  The best fit Gaussian model for the line yields a mean redshift of $z_{[C\, II]}=4.3415 \pm 0.0003$ with $\Delta V = 366$ \kms, reflecting the asymmetry with brighter emission on the blue-shifted side of the line.   The origin of this line asymmetry is unknown at the moment, and future high spectral resolution measurements with a better $S/N$ should reveal whether this asymmetry is also present in CO lines and potentially offer a useful insight into the spatial distribution between the CO and the \C2\ emitting gas.  

The measured \C2\ line integral of $S_{[C\, II]}\Delta V = 13.05 \pm 0.70$ Jy km s$^{-1}$ translates to \C2\ line luminosity of 
$L_{[C\, II]}=1.04\times 10^{-3}\, S_{[C\, II]}\Delta V\, \nu_0 (1+z)^{-1} D_L^2 = 7.8\pm1.1 \times 10^9 L_\odot$.  This is nearly 30 times larger than the line luminosity of both CO transitions detected with the RSR (see Table~\ref{tab:CO}).  To put this in perspective, the \C2\ line luminosity of COSMOS AzTEC-1 is  only a factor of 3 smaller than the {\em total} IR luminosity of an $L^*$ galaxy in the local universe \citep{soifer87,saunders90,yun01}.  Indeed COSMOS AzTEC-1 is extremely luminous in \C2\ line, and this explains why the \C2\ line is detected so easily by the SMA.  Nevertheless, this \C2\ line luminosity is only 0.04 per cent of the total IR and bolometric luminosity (see below), and this ``[C~II] deficiency" is discussed further in detail in \S~\ref{sec:C2}.

Both the \C2\ line and the 345 GHz continuum detected are centered precisely on the position of the \Hubble\ ACS $i$-band source as shown in Figure~\ref{fig:345continuum}.  The total measured 345 GHz continuum flux of $17.8\pm1.4$ mJy is in excellent agreement with the 340 GHz continuum fluxes of $15.6\pm1.1$ mJy previously reported by \citet{younger07}.  This continuum is only marginally resolved by the longest baselines of the SMA with an inferred Gaussian source diameter of only 0.3\arcsec \citep{younger08}.  Both the line and continuum emission is unresolved by the $5.8\arcsec \times 3.4\arcsec$ resolution of the new SMA data, as expected.

\subsection{IR Luminosity and SFR}

The extensive and deep multi-wavelength photometric data readily available in the COSMOS field allowed \citet{smolcic11} to assemble an impressive array of spectral energy distribution (SED) data for their photometric redshift analysis as well as stellar population modeling and bolometric luminosity estimation -- see their Table~1 and Figure~4.   By adding the \Herschel\ SPIRE 250, 350, and 500 \micron\ photometry \citep{smith12} as well as our new 345 GHz continuum measurement (see \S~\ref{sec:C2}), we have fully mapped the infrared peak of the SED as shown in Figure~\ref{fig:SED}, and a more reliable analysis of the dust heating and infrared luminosity is now possible.   

We model and interpret the observed SED using three different commonly used tools: a modified black body model, a starburst SED model by \citet{efstathiou00}, and the GRASIL population synthesis and radiative transfer model \citep{silva98}.  The modified black body model characterizes only the far-IR part of the SED as dust  processed radiation at equilibrium temperature weighted by luminosity.  The two latter models aim to gain further insight into the nature of the luminosity sources by adding assumptions on the source geometry and star formation history.  All these models are highly idealized, however, and these interpretations should be taken in the context of the assumptions adopted.

\subsubsection{Modified Black Body Model}

A common illustrative model for thermal dust emission from astronomical sources is modified black body radiation or ``grey body" radiation.  Following the classical derivation by \citet{hildebrand83} and adopting an emissivity function of the form $Q(\nu)=1 - exp[-(\frac{\nu}{\nu_c})^\beta]$ (so that the emerging spectrum is pure black body spectrum at $\nu \gg  \nu_c$ and $S_\nu \propto \nu^{2+\beta}$ at lower $\nu$) leads to a simple functional form of $S_d(\nu)=\Omega_d B(\nu,T_D) [1-exp[-(\frac{\nu}{\nu_c})^\beta]$, where $\Omega_d$ is the solid angle of the source and $B(\nu,T_D)$ is Planck Function at frequency $\nu$ and dust temperature $T_d$ \citep{yun02}.   

The best fit model that describes the observed SED between 250 and 1000 \micron\ photometry measurements is shown in Figure~\ref{fig:SED}, characterized by dust temperature $T_d=54\pm3$ K and emissivity $\beta=1.6\pm0.2$.   The derived IR luminosities are $L_{IR}=1.4\times 10^{13} L_\odot$ and $L_{FIR}=1.0 \times 10^{13} L_\odot$, where $L_{IR}$ and $L_{FIR}$ are luminosities between $\lambda=8-1000$ \micron\ and $\lambda=40-120$ \micron, respectively.  These luminosity estimates agree well with other estimates discussed below, although they are slightly smaller because this single dust component characterization does not account for the warm dust contribution in the mid-IR region.  The derived dust emissivity $\beta=1.6$ is similar to the commonly adopted value of 1.5 and is known to be somewhat degenerate with $T_d$ in this formulation.  The dust temperature of AzTEC-1 follows the general trend of increasing $T_d$ with $L_{IR}$ reported by recent statistical studies such as by \citet{symeonidis13} and \citet{magnelli12,magnelli14}.  The derived dust temperature of 54 K for AzTEC-1 is higher than the average $\left< T_d \right> \approx 40$ K for $L_{IR}=10^{13} L_\odot$ sources at $z=2$ analyzed in these studies, and their predictions on the redshift evolution of $T_d$ with $L_{IR}$ differ slightly -- sample selection is likely important.  Far-IR data for $z>4$ sources are rare because of the limitations of existing facilities, but our derived dust properties are similar to those of the seven $z>4$ SMGs with $L\sim10^{13}L_\odot$ analyzed by \citet[][$T_d=$ 40--80 K]{huang14}.

\subsubsection{Starburst SED Model by Efstathiou et al. (2000)}

\begin{figure}
\includegraphics[width=0.99\columnwidth]{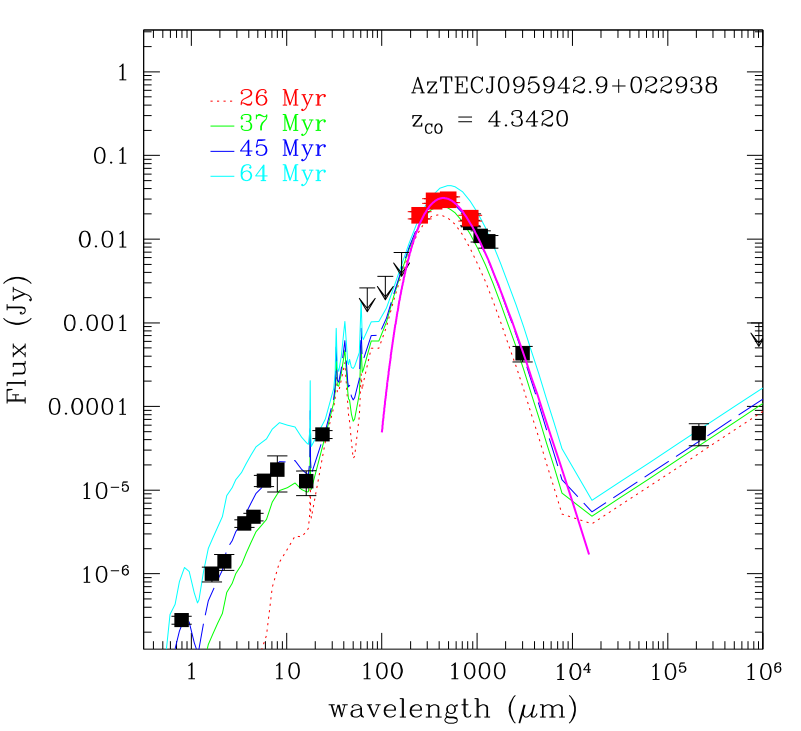}
\caption{A spectral energy distribution (SED) of COSMOS AzTEC-1 from UV to radio wavelengths. Most of the photometry points shown are already summarized in Table~1 by \citet{smolcic11}.  The new 345 GHz SMA photometry and the \Herschel\ SPIRE 250, 350, and 500 \micron\ points (shown in red) are from the published catalog by \citet{smith12} clearly map out the dust peak, allowing us an accurate IR luminosity for the first time.  The best fit modified black body model fitting the far-IR part of the SED is shown in magenta line (see the text for details).   The ``starburst" model SEDs by \citet{efstathiou00} with ages of 26, 37, 45, \& 64 Myr are shown for comparison.}
\label{fig:SED}
\end{figure}

Model SEDs of young stellar clusters embedded in a giant molecular cloud by \citet{efstathiou00} are shown in Figure~\ref{fig:SED}, primarily for illustrative purposes and to compute the IR luminosity of COSMOS AzTEC-1.  Although based on a relatively simple geometry 
and a highly idealized star formation history (a $\tau$-model), these ``starburst" models as well as ``cirrus" models with a lower opacity are shown to be remarkably effective in reproducing the observed SEDs of high redshift ultraluminous infrared galaxies (ULIRGs) and SMGs \citep[see][]{efstathiou09}.   The main impact of increasing starburst age is the build-up of the photospheric emission at wavelengths shorter than 3 \micron\ (with a corresponding increase in  luminosity in the rest frame optical bands) and a systematic shift of the dust peak to a longer wavelength as average opacity decreases and lower mass stars contribute more to the luminosity (i.e., cooler dust temperature).   The success of these relatively simple SED models can be attributed at least in part to the basic fact that the youngest stars dominate the luminosity and the detailed star formation history is largely washed out.  Thus, we expect the IR luminosity and the current star formation rate to be reliable but the mass of the stars produced by the ongoing starburst to be less certain.  The SED with 45 Myr old starburst is in very good agreement with nearly every photometry data in Figure~\ref{fig:SED}, but this time scaling may be meaningful only in the context of this specific model.  
The radio to millimetre wavelength part of the SED is constructed using the well established radio-IR correlation for star forming galaxies as described by \citet{yun02}, and the observed 1.4 GHz radio emission is entirely consistent with the radio and IR luminosity being powered by a pure starburst.

Using the best fit SED (``45 Myr") model shown in Figure~\ref{fig:SED}, we estimate the luminosities of  $L_{IR}=1.5 \times 10^{13} L_\odot$ and $L_{FIR}=9.1 \times 10^{12} L_\odot$, where $L_{IR}$ and $L_{FIR}$ are luminosities between $\lambda=8-1000$ \micron\ and $\lambda=40-120$ \micron, respectively.  The star formation rate derived from $L_{IR}$ using the empirical calibration by \citet[][i.e., $SFR=L_{IR}/(9.4\times 10^9 L_\odot)\, M_\odot /yr$ adjusted for Kroupa IMF]{kennicutt98} is $1596\, M_\odot$/yr.   Assuming $z=4.64$, \citet{smolcic11} estimated $L_{IR}=2.9\times10^{13}L_\odot$, nearly a factor of 2 larger than our estimate when corrected for the new redshift, because they did not have the \Herschel\ photometry to constrain the FIR peak.   In comparison, the current $SFR$ computed by the best fit SED model shown in Figure~\ref{fig:SED} is $880\, M_\odot$/yr, about a factor of 2 smaller than estimated from the IR luminosity.  The smaller current $SFR$ of this model stems from the exponentially decreasing star formation scenario adopted by the model and may not be accurate. 

\subsubsection{GRASIL SED Models \label{sec:GRASIL}}

GRASIL is a population synthesis code which predicts the SED of galaxies from far-UV to radio wavelengths \citep{silva98}.  By allowing a wide range of geometry for gas/dust and stars and a realistic treatment of dust processing as well as a variety of star formation histories and IMFs, GRASIL can model SEDs of a wide range of plausible astrophysical scenarios.  To model the observed SED of COSMOS AzTEC-1, we adopt a Schmidt type law ($SFR(t)=\nu_{Sch}M_{gas}^k$) with an efficiency of $\nu_{Sch} = 0.5$ Gyr$^{-1}$ and an exponent $k = 1$ for the quiescent star formation history. The starburst component is modeled as an exponentially decreasing $SFR(t)$ with an e-folding time $t_b$, observed at different times (``$age_b$") after the outset of the burst.  
Both stellar and gas/dust sources are modeled with a King profile with core radii of $r_*$ and $r_{gas}$.  Two different components of gas and dust are also considered: (a) molecular clouds (MCs) where young stars are forming; and (b) ``diffuse" or ``cirrus" component that surrounds the MCs, old free stars, and exposed new stars with a large filling factor.  A self-consistent radiative transfer calculation is performed to compute the emerging SED.  Uncertainties in the model parameters are estimated from the analyses of 250 Monte Carlo realisations of the input photometry data.

\begin{figure}
\centering
\includegraphics[width=0.99\columnwidth]{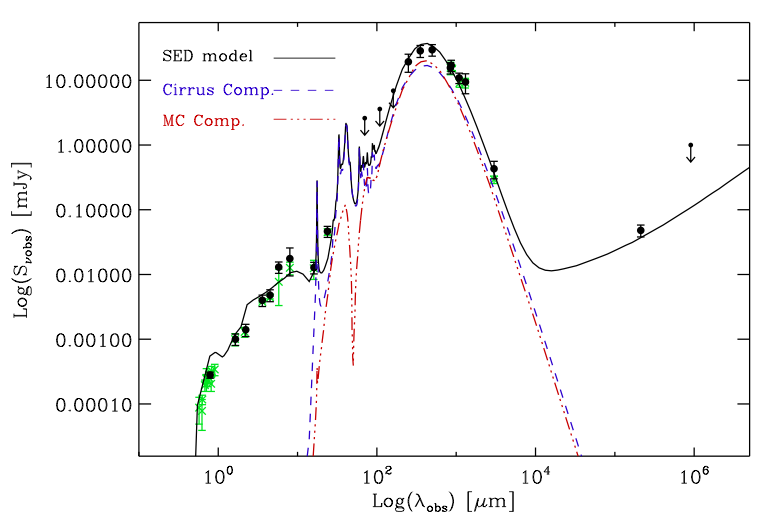}
\caption{Best fit GRASIL starburst SED model for COSMOS AzTEC-1 is shown with a solid line.  In addition to the photometry points shown in Figure~\ref{fig:SED}, additional photometry data in the UV and optical bands compiled by \citet{smolcic11} are also shown [in green].  The observed dust reprocessed (FIR) light originates nearly equally from the diffuse medium (blue dashed line) and the molecular clouds (red dotted-dashed line).}
 \label{fig:grasilsed}
\end{figure}

The best fit GRASIL starburst SED model for COSMOS AzTEC-1 is shown in Figure~\ref{fig:grasilsed} along with the photometry data used to constrain the model.  An acceptable fit could be obtained without any AGN contribution, and the observed SED is consistent with the bulk of the luminosity being powered by a strong starburst,  similar to the majority of local ultra-luminous infrared galaxies \citep[ULIRGs; ][]{vega08}. The starburst component of the best fit model is characterized by an exponentially fading burst with an e-folding time of $t_b=35$ Myr, observed at $age_b=51$ Myr.  At these time scales, the majority of young stars are still embedded within their parent molecular clouds, and most of their luminosity emerges in the IR.  Regardless of any detailed assumptions in the model, the global quantities such as luminosity and $SFR$ should be quite robust.
 The total IR and FIR luminosities derived are $L_{IR}=1.6\pm0.3 \times 10^{13} L_\odot$ and $L_{FIR}=1.2\pm0.2 \times 10^{13} L_\odot$, respectively, with the current star formation rate of $SFR=1320\pm230\, M_\odot$ yr$^{-1}$.  These IR luminosities are in good agreement with those derived in the previous section, and the model $SFR$ is close to $SFR$ derived from the IR luminosity, $1702\pm296\, M_\odot$ yr$^{-1}$.   

There is little evidence to support any AGN activity in AzTEC-1, despite its large luminosity ($L_{IR}>10^{13}L_\odot$).  While the light distribution seen in the \HST\ $i$-band image (Fig.~\ref{fig:345continuum}) is compact, it is clearly resolved with a diameter of 0.3$\arcsec$, similar to other $z>3$ SMGs that are compact and clumpy with $r_e\le 2$ kpc \citep{toft14}.  The photometry data is sparse in the mid-IR  (5-50 \micron) range, and the presence of a heavily obscured AGN cannot be completely ruled out by this modeling.  The red continuum between 0.8-8.0 \micron\ can be interpreted as an indication of a buried AGN \citep{lacy04,stern05}, but it is also the characteristics of a heavily obscured young star clusters, commonly seen among most SMGs \citep{yun08}. The observed radio continuum flux is entirely consistent with the expected supernovae rate (see Fig.~\ref{fig:SED}), and there is no room for any significant AGN contribution in the radio either.

A good spectral coverage from the UV to the radio imposes strong constraints, not only on the global properties of the galaxies but also on other important physical parameters. In the absence of an AGN, the fit to the near-IR  and radio luminosities provides a strong constraint on the star formation rate, while the detailed shape of the SED is affected mainly by the value of the model parameters such as star formation history and extinction.  For instance, the SED shape in the UV range is determined by the geometry between stars and dust while the SED in the optical range can help us to put constraints on the age of the old/intermediate age stellar populations.  The optical depth mainly affects the mid-IR spectral range by varying the MC contribution.   The best fit model core radii of the stellar and gas distributions are $r_*=0.10\pm0.02$ kpc and $r_{gas}=0.95\pm0.42$ kpc, respectively, in agreement with their sizes measured by the \HST\ and \SMA.  The larger extent of gas and dust over the stars ensures an efficient obscuration of the stellar light, and the high mean opacity ($A_V>200$ for the MC component), shapes the observed very red continuum SED in the rest frame optical and near-IR bands, as commonly seen in other high redshift SMGs \citep[e.g.,][]{yun08,yun12}. 

The total stellar mass derived by the GRASIL model $M_*=4.4\pm0.7 \times 10^{11} M_\odot$ is nearly twice as large as the estimate by \citet{smolcic11}, but this estimate depends strongly on the chosen star formation history and thus is not very secure. 
The total gas mass inferred from the GRASIL model is $3.6\pm0.6 \times 10^{11} M_\odot$, and the 39 per cent of this total ($1.4\times 10^{11} M_\odot$) is in the ``dense" (or MC) phase directly fueling the star formation.  As discussed below, deriving the total molecular gas mass from the new CO measurements requires several highly uncertain assumptions, and the gas mass estimate by the GRASIL model is near the high end of the estimates derived using different methods and is close to the gas mass estimated from the dust luminosity using the relation derived by \citet{scoville14}.  The gas mass fraction $f_{gas}\equiv \frac{M_{gas}}{M_{gas}+M_*}$ of 45 per cent is significantly higher than the value derived in nearby galaxies of 10-20 per cent  and is similar to the mass fraction found for the $z\sim2$ submillimetre galaxies \citep[e.g.,][]{tacconi10,geach11}.

About 50 per cent of the IR luminosity arises from the MC component while the other 50 per cent comes from the reprocessed light from free stars in the cirrus component.  The average density of the dense MC component,  $n_{MC} = 7\times 10^5$ cm$^{-3}$, is higher than the critical density of CO (4--3) and CO (5--4) transitions, and this gas is capable of producing fully thermalized emission in these transitions.  The GRASIL model does not constrain the nature of the cirrus component well, but the average density is expected to be 2-3 orders of magnitude lower when scaled by mass and sizes, and the two high $J$ CO transitions are expected to be sub-thermally excited in this diffuse component.  

The best fit GRASIL mode for COSMOS AzTEC-1 is remarkably similar to the best fit starburst SED models by \citet{efstathiou00}, as the gross shapes of the SED are driven largely by a young starburst embedded in dense gas clouds in both models.  The best fit GRASIL model also requires a starburst history remarkably similar to the Efstathiou models discussed above, and this offers a plausible explanation for the apparent success of the Efstathiou models despite its simplicity.  With a more sophisticated parameterization, the GRASIL model can offer a more nuanced physical insight into the gas, dust, and stellar properties.

\section{Discussion}

\subsection{Gas Mass and Nature of the Molecular Gas Fueling the Luminosity \label{sec:mass}}

A standard practice for deriving total molecular gas mass from a redshifted CO line measurement is to adopt a standard line ratio between different rotational transitions of CO to estimate the luminosity of  the CO (1--0) transition $L'_{CO(1-0)}$ and then to translate this line luminosity to molecular gas mass assuming an ``$\alpha_{CO}$" conversion factor \citep[see a review by][]{carilli13}.  The Table~2 by \citet{carilli13} gives the average CO line ratios for SMGs as $L'_{CO(4-3)}/L'_{CO(1-0)}=0.46$ and $L'_{CO(5-4)}/L'_{CO(1-0)}=0.39$.  Using these ratios, the measured CO line luminosities $L'_{CO(4-3)}$ and $L'_{CO(5-4)}$ in Table~\ref{tab:CO} can be translated to $L'_{CO(1-0)}$ of $1.7\pm0.2 \times10^{11}$ and $1.1\pm0.2 \times10^{11}$ K km s$^{-1}$ pc$^{-2}$, respectively.  Taking CO(4--3) line luminosity (which requires a smaller correction to CO(1--0) line luminosity) and a ``ULIRG" conversion factor of $0.8 M_\odot [{\rm K\, km s^{-1}\, pc^{-2}}]^{-1}$, we derive a total molecular gas mass of $1.4\pm0.2 \times 10^{11} M_\odot$.   Using a Galactic conversion factor of $\alpha_{CO} \equiv M_{H2}/L'_{CO(1-0)}$ of $\sim4 M_\odot [{\rm K\, km s^{-1}\, pc^{-2}}]^{-1}$ yields a 5 times larger total molecular gas mass, $6.8\pm0.8 \times 10^{11} M_\odot$. 

Since we have spectral measurements of multiple CO lines, \C2\ line, nearly fully mapped SED, and spatially resolved dust continuum distribution, we should be able to probe the gas properties of AzTEC-1 beyond simply adopting a highly uncertain and somewhat arbitrary ``average" calibration.  In this section, we explore several different methods for estimating gas mass, including dynamical mass analysis, radiative transfer analysis, and dust continuum measurements, in order to obtain a better handle on the gas mass and excitation conditions.

\subsubsection{Gas Mass from Dynamical Mass \label{sec:dyn_mass}}

There is a growing awareness that the CO-to-H$_2$ conversion factor is not a single value but a quantity dependent on several different factors, such as metallicity, density, temperature, and non-gravitational pressure \citep[see a review by][]{carilli13}.  Since CO is a highly optically thick transition, metallicity is important mainly for low metallicity systems and should not be an important factor for SMGs -- they are selected by their bright dust continuum and are often bright in CO, \C2, and other transitions that require near solar abundance of metals.  The main reason why the ``ULIRG" conversion factor is often favored for SMGs is that the observed line width may include significant other pressure contributions such as stellar mass and stronger turbulence, as well as a higher radiation field resulting from a high density of young stars, similar to the central kpc of nearby ULIRGs.  \citet{downes98} derived the ULIRG conversion factor, $\sim$5 times smaller than the MW value {\em on average}, by constructing a dynamical model of spatially resolved CO emission in 10 nearby ULIRGs and by running a full radiative transfer calculation, also taking into account the mass contributions by the new and old stars.  Because of poor spatial resolution of nearly all existing molecular line imaging data, few examples of such a dynamical mass analysis exist for SMGs.  In one of the best studies of high redshift SMGs using this approach, \citet{hodge12} have derived $\alpha_{CO} = 1.1\pm0.6 M_\odot [{\rm K\, km s^{-1}\, pc^{-2}}]^{-1}$ for the $z=4.05$ SMG GN20, which has a rather large ($14\pm4$ kpc diameter) CO disk with a dynamical mass of $M_{dyn}= 5.4\pm2.4 \times 10^{11}M_\odot$, stellar mass of $M_*= 2.3\pm2.4 \times 10^{11}M_\odot$, and a total gas mass of $M_{H2}= 1.8\pm0.7 \times 10^{11}M_\odot$.  

We do not have a spatially resolved CO map of COSMOS AzTEC-1, but adopting its 890 \micron\ source size (also the optical source size\footnote{This stellar source size is estimated from the spatially resolved \HST\ $i$-band image.  \citet{toft14} report an upper limit of $r_{e,NIR}<2.6$ kpc based on their analysis of the UltraVista near-IR images.}) of 0.3\arcsec\ (2.1 kpc) as the diameter of the CO emitting molecular gas disk and 1/2 of the observed FWZI \C2\ line width of 219 km s$^{-1}$ (see \S~\ref{sec:C2}) as the rotation velocity $v_c \times (sin~i)$ where $i$ is the disk inclination, a dynamical mass can be computed as $M_{dyn}=1.1(sin~i)^{-2}\times10^{10}M_\odot$. To reconcile this value with its stellar mass \citep[$1.5\pm0.2 \times 10^{11}M_\odot$,][or $4.4 \times 10^{11}M_\odot$ according to our GRASIL model]{smolcic11} as well as the molecular gas mass derived using the ULIRG conversion factor ($M_{H2} \sim 10^{11} M_\odot$), this disk has to be highly inclined, $i \lesssim 12^\circ$.  The gas disk traced in \C2\ line in the $z=4.76$ SMG ALESS 73.1 is 2.2 times larger than the continuum \citep{debreuck14}, and the dynamical mass should double if the gas disk emitting \C2\ in AzTEC-1 is also twice as large.  However,  the required inclination changes only slightly, $i \lesssim 17^\circ$. 
This situation is similar to the nearby ULIRG Mrk~231, which has a 450 pc radius molecular gas disk with an inclination of $i\sim10^\circ$ \citep{bryant96,downes98}.  This nearly face-on geometry also offers a natural explanation as why COSMOS AzTEC-1 is exceptionally bright in the optical bands.  However, its dynamical mass is highly uncertain because of the unknown and small inclination angle, and unfortunately this approach does not offer much useful insight into the total gas mass and the CO-to-H$_2$ conversion factor for COSMOS AzTEC-1.  

\subsubsection{Optically Thin CO: a Low Mass Limit}

Approaching the conversion factor problem from an entirely theoretical point of view, a lower limit to the total gas mass can be derived by considering an optically thin CO case.  \citet{bryant96} have offered a rather detailed derivations of gas mass determination based on spatially resolved CO observations and dynamical modeling in their effort to determine the gas mass for Mrk~231.  Their derivation shows that the optically thin limit conversion factor should be  $\alpha_{CO,min}=0.20 (X_{CO}/10^{-4})^{-1}$, where $X_{CO}$ is the CO abundance.   The minimum gas mass for COSMOS AzTEC-1 derived from this relation is $\sim 3\times 10^{10} M_\odot$ if the gas excitation is typical of SMGs.  The minimum gas mass required increases to $M_{H_2}\sim 10^{11} M_\odot$ if the excitation temperature is lower as indicated by the measured CO(4--3)/CO(5--4) line ratio (see below).  These values can be understood better in the context of gas excitation provided the radiative transfer models discussed below.

\subsubsection{Gas Mass from Radiative Transfer Modeling}

 The measured $L'_{CO(4-3)}/L'_{CO(5-4)}$  line ratio of $1.78\pm0.35$ for COSMOS AzTEC-1 is larger than the average SMG ratio of $0.46/0.39=1.18$ reported in the recent review by \citet{carilli13} and is closer to the average Milky Way GMC value of 2.1 -- see their Table~2.  Although omitted by the Carilli \& Walter review, the scatter in the SMG line ratios found in the literature is rather large because many published CO measurements have low $S/N$ ratios.\footnote{For example, in one of the largest recent studies of multiple CO transitions by \citet{bothwell13}, 18 out of 32 (56\%) CO line measurements used for analyzing the SMG line ratios have $S/N<5$ -- see their Table~5. }  In contrast, the line ratio for AzTEC-1 is quite secure not only because each line is detected with $S/N>7$ individually, but also because the two CO lines are measured simultaneously using the RSR, removing any potentially large systematic uncertainties arising from  utilizing measurements taken at different times using  different instruments, often on different telescopes.   

We explore the mass and physical conditions of the gas producing the observed CO emission in COSMOS AzTEC-1 by examining a grid of models covering a range of density $n$, kinetic temperature $T_{kin}$, and the CO column density $N_{CO}$ that can reproduce the measured CO line intensities and line ratios.  If the gas was optically thin, then only density, temperature and total number of CO molecules are needed to model the observed emission.  However, it is more likely that the CO emission is optically thick, and this impacts not only the escape of CO photons from the molecular gas, but also the excitation of CO column trapping.  The optical depth of the CO emission is determined by the ratio of the CO column density to the line width $\Delta V$, and we have run models for three cases in which $N_{CO}/\Delta V = 2\times 10^{14}$ cm$^{-2}$ per km s$^{-1}$, $6\times 10^{16}$ cm$^{-2}$ per km s$^{-1}$, and $2\times 10^{17}$ cm$^{-2}$ per km s$^{-1}$.  In the first case the CO emission in the $J=4-3$ line is optically thin ($\tau_{CO} \ll 1$).  In the second case, the optical depth in this line is modest ($\tau_{CO} \approx 5$), and in the final case the optical depth is large ($\tau_{CO} \approx 20$).
The excitation by the Cosmic Microwave Background (CMB) with $T_{CMB}=14.6$ K at $z=4.342$ is explicitly included in these calculations.  

\begin{table*}
 \caption{Summary of the RADEX models with acceptable solutions}
 \label{tab:RADEX}
 \begin{tabular}{@{}cccccc@{}}
  \hline
  \hline
  n & $T_{kin}$  & $I_{CO1-0}/I_{CO4-3}$ & $I_{CO1-0}/I_{CO5-4}$  & $I_{CO4-3}/N_{CO}$ & $M_{H2}$ \\
  (cm$^{-3}$) &  (K) & & & (K km s$^{-1}$ cm$^2$) & ($M_\odot$)  \\
 \hline
 \hline
 \multicolumn{6}{c}{$N_{CO}/\Delta V = 2\times10^{14}$ cm$^{-2}$ per km s$^{-1}$ (optically thin CO cases)}\\
 \hline
 $1\times10^3$ & 500 & 1.28 & 2.30 & $6.9\times10^{-16}$ & $1.8\times10^{10}$ \\
 $3\times10^3$ &  200  & 0.79   &     1.50    &    $9.6\times10^{-16}$  &  $1.3\times10^{10}$ \\
 $1\times10^4$ &     100     &   0.36   &     0.63    &    $1.4\times10^{-15}$  &  $8.9\times10^9$ \\
 $3\times10^4$ &        50     &   0.30   &     0.54    &    $1.4\times10^{-15}$  &  $8.9\times10^9$\\
 $1\times10^5$ &       33    &    0.31   &     0.55    &    $1.2\times10^{-15}$  &  $1.0\times10^{10}$ \\
 $3\times10^5$ &       28    &    0.33     &   0.58     &   $1.0\times10^{-15}$   & $1.2\times10^{10}$ \\
 $1\times10^6$ &       25     &   0.37     &   0.68    &    $8.5\times10^{-16}$  &  $1.5\times10^{10}$ \\
 \hline
 \hline
 \multicolumn{6}{c}{$N_{CO}/\Delta V = 6\times10^{16}$ cm$^{-2}$ per km s$^{-1}$ (modest CO optical depth cases)}\\
 \hline
$1\times10^3$     &   120    &   2.6    &        4.8    &    $1.5\times10^{-16}$   & $8\times10^{10}$ \\
$3\times10^3$    &     80    &    1.7     &       2.9     &   $2.4\times10^{-16}$   & $5\times10^{10}$ \\
$1\times10^4$     &    35    &    1.4      &      2.4    &    $1.8\times10^{-16}$  &  $7\times10^{10}$\\
$2\times10^4$	&	 25	  &  1.3	     &  2.2	   &   $1.1\times10^{-16}$	   &  $1\times10^{11}$ \\
$3\times10^4 $    &    20    &    1.3     &       2.3     &   $6.3\times10^{-17}$  & $2\times10^{11}$\\
$1\times10^5$     &    15     &   1.3     &       2.3    &    $5.0\times10^{-18}$   & $2.5\times10^{12}$\\
$3\times10^5$     &       \multicolumn{5}{c}{No Solution} \\
\hline
 \hline
 \multicolumn{6}{c}{$N_{CO}/\Delta V = 2\times10^{17}$ cm$^{-2}$ per km s$^{-1}$ (optically thick CO cases)}\\
 \hline
$1\times10^3$    &     85      &      2.1     &       3.6  &  $7.9\times10^{-17}$  &  $1.6\times10^{11}$\\
$3\times10^3$     &    35     &       1.7     &       3.1   & $5.2\times10^{-17 }$ &  $2.4\times10^{11}$\\
$6\times10^3$	&	 25	 &      1.5	 &      2.6	& $3.4\times10^{-17}$  &	 $3.7\times10^{11}$\\
$1\times10^4$      &   20      &      1.4     &       2.5   & $1.9\times10^{-17}$  &  $6.6\times10^{11}$\\
$3\times10^4$     &       \multicolumn{5}{c}{No Solution} \\
\hline
\hline
 \end{tabular}
%\medskip
%\noindent An intrinsic line width of $\Delta V=5$ km s$^{-1}$ is assumed.
\end{table*}

Models producing acceptable solutions are summarized in Table~\ref{tab:RADEX}. The gas density and temperature that satisfy the observed line ratio are given
along with the ratios of the CO (1--0) line intensity relative to the CO (4--3) and CO (5--4) lines and the integrated intensity in the CO (4--3) line divided by the CO column density (in units of K km s$^{-1}$ per cm$^{-2}$).  If we make the assumption that the abundance of CO relative to molecular hydrogen is $X_{CO}=10^{-4}$, then we can determine the amount of mass required to produce the observed $L'_{CO}$ in the CO (4--3) line.  The last column gives the required gas mass for that model. 

At a CO column density per unit line width of $N_{CO}/\Delta V = 2\times10^{14}$ cm$^{-2}$ per km s$^{-1}$ where all CO transitions are optically thin, an acceptable solution is found over a broad range of temperature.  These optically thin cases require the least amount of gas mass to reproduce the observations among the models examined, but one still needs at least of order $10^{10}M_\odot$ of gas to produce the observed CO line emission, unless CO is more abundant than it is in the Milky Way.   A comparison with the optically thin, low limiting mass calculation based on the derivation by \citet{bryant96} above suggests that the $L'_{CO(1-0)}$ estimates derived using the average SMG CO line ratios taken from \citet{carilli13} are 2-3 times too large, or the CO emitting gas has to be cold ($\lesssim25$ K).

For the modest to high optical depth cases with $N_{CO}/\Delta V = 6\times10^{16}$ cm$^{-2}$ per km s$^{-1}$ and $2\times10^{17}$ cm$^{-2}$ per km s$^{-1}$, acceptable solutions exist only for a narrower range of density and temperature: $10^3 \lesssim n \lesssim 10^5$ and $15 \lesssim T_{kin} \lesssim 120$ K.  The required gas mass is at least $5\times10^{10} M_\odot$ in these cases, and more likely the range of mass is $(1-7) \times 10^{11}M_\odot$ (a higher mass for lower $T_{kin}$ and higher $N_{CO}/ \Delta V$), comparable or exceeding the estimated total stellar mass.  If the mean gas density for the star forming gas (the ``MC" component) is $\gtrsim 10^5$ cm$^{-3}$ as suggested by the GRASIL model (see \S~\ref{sec:GRASIL}), then there is a very small range of parameter space in density and temperature where an acceptable solution exists, and these solutions favor cold gas temperature ($\lesssim25$ K) with a large total mass ($M_{H2}>(2-7)\times10^{11} M_\odot$).  
An important caveat for this radiative transfer calculation is its assumption of a single component gas.  If  more than one phase gas with vastly different excitation conditions (e.g., ``cirrus" and ``MC") contribute {\em significantly} to the observed CO line intensities and line ratios, then a single component analysis such as presented here may lead to a misleading result.  
While a solution satisfying the observed $L'_{CO(4-3)}/L'_{CO(5-4)}$  line ratio can be found in a fairly broad range of excitation conditions and CO column density, an important merit of the models summarized Table~\ref{tab:RADEX} is the clear prediction they make on the CO (1--0) line intensity, that the optically thick cases should produce five times stronger CO (1--0) line than the optically thin cases (for $T_{kin}\le200$ K).  Therefore, future CO (1--0) line measurements should yield an effective discrimination between these limiting cases and the gas properties when combined with these measurements.

We briefly explored using \C2\ line intensity to gain further constraints on the gas excitation and mass, but we found this even more problematic.  The critical density for excitation of \C2\ line is much lower ($n_{H2}\ge10^3$ cm$^{-3}$) than those of the CO transitions we measured, making this analysis more susceptible to the likely presence of multiple gas phases.  Furthermore, the \C2\ line can originate from both neutral and ionized gas \citep[with critical electron density of $n_e=10-100$ cm$^{-3}$, see Table~2 by][and discussions below]{goldsmith12}, and we do not have much confidence in making the assumption that \C2\ and CO lines arise from the same gas.
 
\subsubsection{Gas Mass from Dust Continuum}

The molecular gas mass derived from these high $J$ rotational transitions are likely lower limits since they may be sub-thermally excited compared with the CO (1--0) transition.  One way to check this is to compare the total gas mass derived from the Rayleigh-Jeans (RJ) part of the dust spectrum as proposed by \citet{scoville14}.  Given the uncertainties in excitation and conversion factor for CO, Scoville et al. have argued that dust mass derived from the RJ part of the dust spectrum and adopting a gas-to-dust ratio is more robust than an estimate based on high $J$ CO line luminosity.   The Eq.~12 by Scoville et al. can be rewritten as
\begin{equation}
M_{ISM} =\frac{1.2\times10^{10}}{(1+z)^{4.8}}[\frac{\Gamma_{RJ}}{\Gamma_0}]^{-1}[\frac{S_\nu}{\rm mJy}][\frac{\nu_{}}{353GHz}]^{-3.8}[\frac{D_L}{\rm Gpc}]^2 M_\odot
\end{equation}
where $S_\nu$ is the observed dust continuum in mJy, $D_L$ is luminosity distance in Gpc, and $\frac{\Gamma_{RJ}}{\Gamma_0}$ is the RJ correction factor (see their Fig.~2).  Using this relation, the 345 GHz continuum measurement from the SMA (see \S~\ref{sec:C2}) can then be translated to a total ``ISM mass" of $M_{ISM}=(7.4\pm1.3)\times 10^{11} M_\odot$ for $T_d=25$ K and $(3.6\pm0.7)\times 10^{11} M_\odot$ for $T_d=35$ K.    
As noted by Scoville et al., estimating dust temperature from the measured dust peak may be a mistake since the observed SED is indicative of luminosity-weighted (rather than mass-weighted) measure of dust temperature.  And the resulting smaller gas mass derived for the higher dust temperature is only a lower limit.  The error in gas mass depends only linearly with any error in dust temperature, and the resulting $\sim$50 per cent uncertainty due to poorly constrained dust temperature still leads to a better gas mass estimate than the estimates from the CO luminosity, which is fraught with a wide range of substantial and systematic uncertainties.   It is notable that the total gas mass derived from the measured dust continuum is among the largest estimates obtained by the different methods, but it agrees well with the estimate by the GRASIL model (see \S~\ref{sec:GRASIL}) and the gas masses required to produce the observed CO line ratio in the optically thick cases (see Table~\ref{tab:RADEX}).   

\subsubsection{Summary of Gas Mass Estimation and Broader Implications}

A detailed review of the process of converting the measured CO (4--3) and (5--4) line luminosity to a total gas mass demonstrates several assumptions one has to make, particularly when using empirical calibrations.   The measured line ratio between these two transitions for COSMOS AzTEC-1 is larger than the average ratio reported for a sample of SMGs by \citet{carilli13}, and the resulting uncertainty in estimating CO (1--0) line luminosity is large (a factor of 2 to 3).   The CO-to-H$_2$ conversion factor is also uncertain by at least a factor of 2 or more, but the nominal total molecular gas mass based on the CO (4--3) line luminosity is $1.4\times 10^{11} M_\odot$.   The non-LTE radiative transfer calculations have yielded a range of acceptable solutions summarized in Table~\ref{tab:RADEX}, with likely molecular gas mass in the range of $(1-7)\times 10^{11}M_\odot$ for the modest to high optical depth cases and with a minimum (optically thin) limit of $\sim 10^{10}M_\odot$.  The total ISM mass estimate based on dust continuum \citep{scoville14} favors the upper end of these estimates, $(4-7)\times 10^{11}M_\odot$ while the empirical calibration that yields a total gas mass of $\sim 2\times 10^{11}M_\odot$ from the measured CO (4--3) line luminosity is at the low end of these different estimates.

Two interesting outcomes from these analysis  deserve additional comments.  Firstly, the gas mass analysis using the dynamical mass and stellar mass estimates has revealed that the geometry of the gas disk has to be nearly face-on, and consequently the poorly constrained dynamical mass prevents us from deriving a meaningful gas mass estimate.  This calculation also sheds  an interesting insight in that such a nearly face-on geometry with minimum dust obscuration would naturally explain why the stellar component of the host galaxy is seen only modestly obscured in the rest frame UV light ($A_V\lesssim 3.5$), unlike most other SMGs with similarly high IR luminosity $\gtrsim10^{13}L_\odot$.   Secondly, the observed CO (4--3) to (5--4) line ratio is lower than the typical value for SMGs and is closer to the Milky Way value.  The radiative transfer models for gas density and temperature characteristic of the MW star forming dense cores ($n\sim 10^{4}$ cm$^{-3}$ and $T\sim 25$ K) require a total gas mass of $(2-4)\times 10^{11}M_\odot$, which is more in line with the gas mass estimate from the dust continuum and the gas mass estimate based on the MW value for $\alpha_{CO}$ ($\sim 5\times 10^{11}M_\odot$).  The range of excitation conditions that can reproduce the observed line ratios and intensities are uncomfortably narrow, however, and this may indicate that the observed CO lines include significant contributions from more than one component of molecular ISM present in this galaxy \citep[e.g.,][]{harris10}, as also suggested by the GRASIL analysis.  

Given the range of gas mass estimates and current star formation rate, the gas depletion time for COSMOS AzTEC-1 is about 200 Myr, with about a factor of 2 overall uncertainty.  This means COSMOS AzTEC-1 will exhaust its gas reserve and will shutoff its star formation activity by $z\approx4$ even without any negative feedback, unless gas continues to flow in at a rate matching the star formation rate, $\dot{M} \approx 10^3 M_\odot$ yr$^{-1}$.   Some gas recycling can extend this time by about 50 per cent, but the stellar feedback process is expected to do more than compensating for this effect.  The stellar mass doubling time $\tau_* \equiv M_*/SFR$ for COSMOS AzTEC-1 is also about 200 Myr, and its substantial stellar mass could have been plausibly built up {\em entirely} during the current episode of starburst \citep[see][]{yun12}.  In such a scenario, the starburst activity would have started around $z\approx5$, ending with a $M_* \gtrsim 6\times 10^{11} M_\odot$ stellar galaxy with $\sim2$ kpc diameter by $z\approx4$, similar to the massive quiescent galaxies reported by \citet{whitaker13}, \citet{straatman14}, and others.

\subsection{\C2/FIR Ratio and High Radiation Field   \label{sec:C2}}

\begin{figure}
\includegraphics[width=0.95\columnwidth]{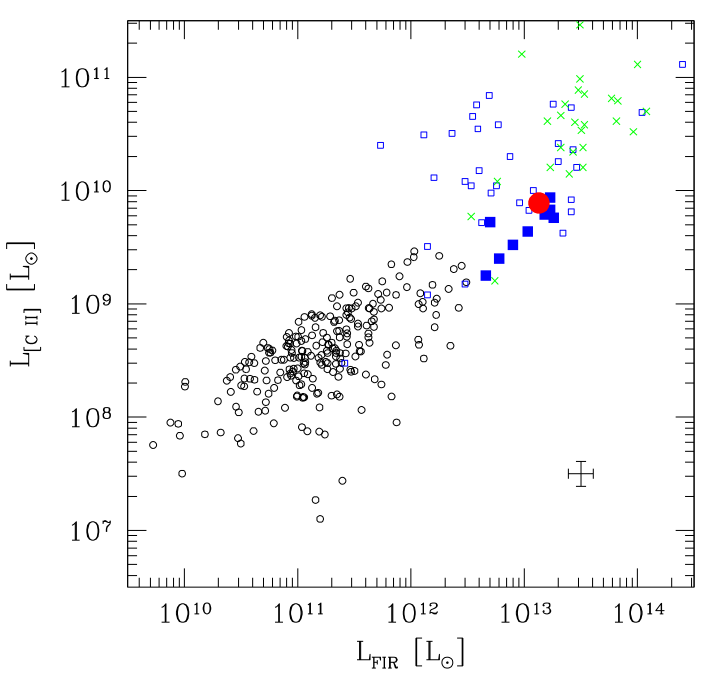}
\caption{$L_{[C\, II]}$ as a function of far-IR luminosity $L_{FIR}$. Empty circles are the GOALS sample of local LIRGs and ULIRGs \citep{diaz13}.  
Empty squares and crosses are [C II] measurements for $z>1$ sources from the literature \citep{walter09,hailey10,stacey10,wagg10,cox11,valtchanov11,gallerani12,swinbank12,venemans12,walter12,carniani13,george13,wang13,magdis14,rawle14,riechers13,willott13,riechers14,brisbin15,gullberg15,schaerer15} -- crosses are strongly lensed sources while squares may also be lensed sources.  COSMOS AzTEC-1 is shown as a large filled circle and extends the trend of ``\C2\ deficiency" to $L_{FIR}\ge 10^{13}L_\odot$. Filled squares are high redshift sources with spatially resolved \C2\ and FIR distribution: COSMOS AzTEC-3 \citep{riechers14}, HDF~850.1 \citep{neri14}, BR1202$-$0725 A\&B \citep{carniani13}, ALESS~73.1\citep{debreuck14}, and four $z=6$ QSOs imaged using ALMA \citep{wang13}.  A typical error bar is shown on the bottom right corner.}
\label{fig:LC2vsLFIR}
\end{figure}

The 158 \micron\ \C2\ line is an important coolant of the neutral ISM and thus is a bright tracer of star formation in galaxies, typically accounting for 0.1-1 per cent of IR luminosity \citep{madden93,malhotra01,stacey10}.   Because \C2\ emission can be produced by different gas phases with a wide range of physical conditions, interpreting \C2\  emission is difficult \citep[see a recent review by][]{goldsmith12}.   A broad correlation is seen between observed \C2\ emission and other tracers of star formation \citep[e.g.,][also see Fig.~\ref{fig:LC2vsLFIR}]{boselli02,delooze11}.  However, interpreting observed \C2\ line luminosity in terms of a particular physical process, such as a tracer of SFR, is problematic because  \C2\ emission arises from a variety of different excitation mechanisms, in both ionized and neutral phase.   Also, star forming galaxies observed in \C2\ show a factor of 100 or more spread in the $L_{[CII]}/L_{FIR}$ ratio, which is correlated with IR luminosity and dust temperature \citep[so-called ``\C2\ deficiency", ][]{malhotra97,luhman98}.  

\begin{figure}
\includegraphics[width=0.95\columnwidth]{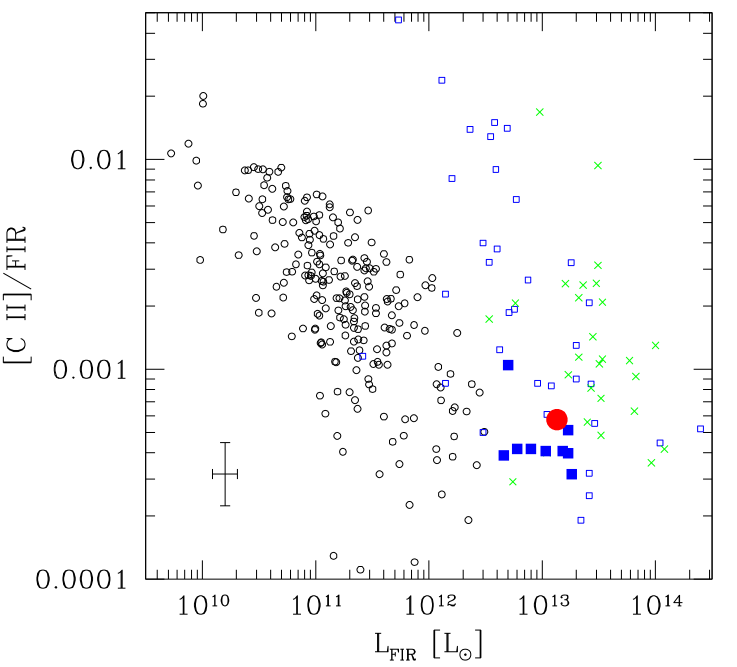}
\caption{\C2/FIR ratio as a function of the FIR luminosity.  Many of the $z>1$ sources are lensed, and their luminosities are not corrected for lensing because the magnification factor is not always known.   All symbols are identical to those in Figure~\ref{fig:LC2vsLFIR}.  A typical error bar is shown on the bottom left corner.}
\label{fig:C2vsFIR}
\end{figure}

The measured \C2\ line luminosity of COSMOS AzTEC-1 is  significantly higher than those of the Great Observatories All-sky LIRG Survey (GOALS) sample of 241 luminous infrared galaxies studied by \citet{diaz13} using the {\it Herschel Space Observatory} (see Figure~\ref{fig:LC2vsLFIR}).  Also shown are a collection of \C2\ line sources at $z>1$ from the literature, and they extends the observed broad correlation to $L_{FIR}>10^{13}L_\odot$.  The [C II]/FIR ratio, shown in Figure~\ref{fig:C2vsFIR}, reveals that the measured $L_{[C\, II]}/L_{FIR}$ ratio of $6.5\times 10^{-4}$  for AzTEC-1 is among the lowest measured and extends the \C2\ deficiency to $L_{FIR}\ge 10^{13} L_\odot$. 

While most of the GOALS sample LIRGs and ULIRGs form a broad trend with a decreasing $L_{[C\, II]}/L_{FIR}$ ratio with increasing $L_{FIR}$, the $z>1$ \C2\ sources show a much larger scatter.   These high redshift systems simply being a scaled up versions of the local star forming galaxies is a commonly offered explanation \citep[e.g.,][]{stacey10,brisbin15}.  At least 1/2 of the sources detected in \C2\ thus far are strongly lensed systems (shown as crosses in Fig.~\ref{fig:C2vsFIR}) found by the South Pole Telescope \citep[SPT; ][]{gullberg15} and \Herschel\ \citep{cox11,valtchanov11,riechers13,magdis14,rawle14}, and many should fall along the local LIRG/ULIRG relation when corrected for magnification, as shown by a detailed study of a $z=2.013$ lensed \C2\ source by \citet{schaerer15}.  Determining whether the larger scatter associated with the remaining $z>1$ sources can be accounted by lensing will require future detailed follow-up studies of the individual sources.

Along with the $z=5.3$ SMG COSMOS AzTEC-3 \citep{riechers14} and the $z=5.2$ SMG HDF~850.1 \citep{neri14}, AzTEC-1 has the one of the smallest $L_{[C\, II]}/L_{FIR}$ ratio in Figure~\ref{fig:C2vsFIR}, clustered together with 7 IR luminous galaxies hosting an optical QSO at $z>4$ imaged in \C2\ and continuum by ALMA (shown as filled squares).  One possible explanation for their extremely low $L_{[C\, II]}/L_{FIR}$ ratio is the elevated dust-obscured AGN contribution to the FIR luminosity, but this explanation is not supported by systematic studies of large samples of IR luminous galaxies conducted using \Herschel.  By examining the GOALS LIRGs with and without AGN activity (identified through \Spitzer\ mid-IR spectroscopy), \citet{diaz13} have shown that the correlation between $L_{[C\, II]}$ and $L_{FIR}$ and the \C2\ deficiency is an intrinsic property of the star formation, and there is no need to invoke AGN activity to explain it (at least at levels of $L_{[C\, II]}/L_{FIR}>10^{-3}$).  A study of 154 intermediate redshift ($\left< z \right>\sim 0.15$) 24 $\mu$-m-selected galaxies by \citet{magdis13} and a study of 130 mid- and far-IR selected galaxies by \citet{sargsyan14} also drew a similar conclusion.   

\begin{figure}
\includegraphics[width=0.95\columnwidth]{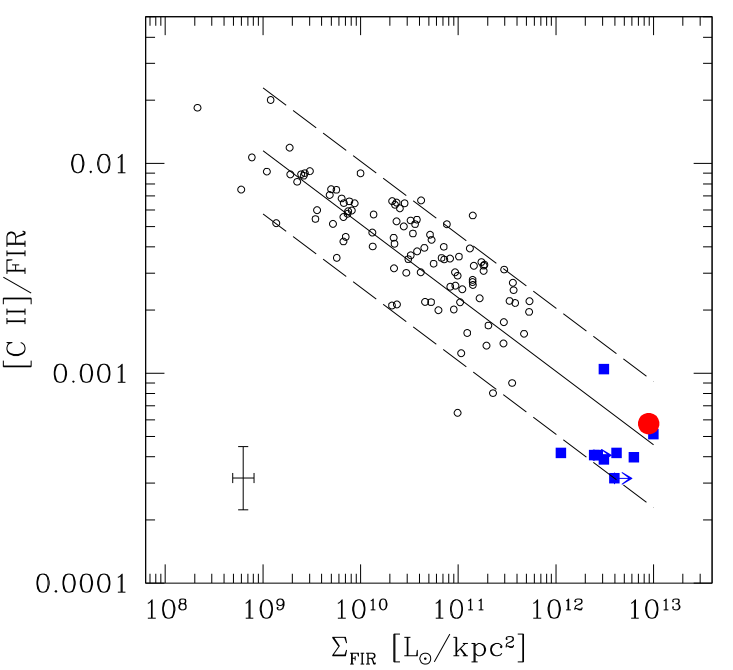}
\caption{\C2/FIR ratio as a function of FIR surface density.  Only the sources with a resolved mid-IR or far-IR sizes are included.  All symbols are identical to those in Figure~\ref{fig:LC2vsLFIR}. A typical error bar is shown on the bottom left corner.}
\label{fig:C2vsSIR}
\end{figure}

The compact source sizes of AzTEC-1 and other sources revealed by high resolution continuum imaging using the SMA and ALMA suggests the high intensity of the infrared radiation field may offer an important clue to the \C2\ deficiency.
Possible explanations for the \C2\ deficiency include: (1) self-absorption; (2) saturation of the \C2\ line due to high gas column density; (3) decreased photoelectric heating in high UV radiation field; and (4) high dust-to-gas opacity caused by an increase of the average ionization parameter \citep[see reviews by][]{malhotra01,diaz13}.   Citing a clear trend for LIRGs with deeper 9.7 \micron\ silicate strengths, higher mid-IR luminosity surface densities ($\Sigma_{MIR}$), smaller fractions of extended emission, and higher specific star formation rates (SSFRs) to display a greater \C2\ deficiency, \citet{diaz13} have concluded that the dust responsible for these correlations must be directly linked to the process driving the observed \C2\ deficiency.  They have also found the correlation becoming much tighter when the FIR luminosity is normalized by mid-IR source size (i.e., surface density $\Sigma_{FIR}$), independent of the nature of the powering source.  As shown in Figure~\ref{fig:C2vsSIR}, COSMOS AzTEC-1 and other high redshift \C2\ sources with spatially resolved continuum sizes  follow the same tight correlation defined by the local LIRGs and extend this correlation by one order of magnitude larger in $\Sigma_{FIR}$.  In addition to extending this correlation to a higher luminosity density, this comparison also further supports the proposed scenario that the compactness of the active region and the resulting higher intensity of the infrared radiation field dictates the \C2\ deficiency.  In an earlier modeling study using the spectral synthesis code CLOUDY, \citet{gracia11} have shown that {\em all} far-IR fine structure lines, regardless of their origin in the ionized or neutral phase of the ISM, show a deficit with increasing $L_{FIR}/M_{H2}$ ratio, and they further conclude that this deficiency is driven by the increased ionization parameter.  This is an extremely interesting prediction that should be tested further using future observations of other far-IR fine structure lines.  Determining through a high resolution imaging study whether the unlensed $L_{FIR}>10^{13}L_\odot$  \C2\ sources with $L_{[C\, II]}/L_{FIR} \gg 10^{-3}$ follow the narrow trend seen in Figure~\ref{fig:C2vsSIR} is another important test for this \C2\ deficiency scenario.

\section{Conclusions}

We report the first successful spectroscopic redshift determination of COSMOS AzTEC-1 obtained with a clear detection of the redshift CO (4-3) and CO (5-4) lines using the Redshift Search Receiver on the Large Millimeter Telescope and the confirmation of the CO redshift through the detection of the redshifted 158 \micron\  \C2\ line using the Submillimeter Array.   Utilizing the newly measured redshift and CO and \C2\ line intensities, we have explored the gas mass and physical conditions of the gas fueling the enormous luminosity associated with this $z=4.342$ SMG.  

The RSR spectrum of COSMOS AzTEC-1 (Figure~\ref{fig:CO}) has two emission lines clearly above the noise level ($\ge7\sigma$) at 86.31 GHz and 107.87 GHz, and they are identified as redshifted CO (4-3) and (5-4) lines at $z=4.3420 \pm 0.0004$, respectively.  This conclusion is supported by its over-all SED as well as by the radio-millimetric spectral index analysis by \citet{carilli99}.  A detailed discussion of the unique redshift determination is presented in the Appendix section.  The derived CO redshift of $z=4.3420$ is slightly lower than the photometric redshift derived by \citet{smolcic11} using the rest frame UV and optical photometry data and is outside the failed previous blind CO searches by \citet{smolcic11} and \citet{iono12}.  This successful redshift determination after nearly 10 years of effort demonstrates the power of the ultra-wideband spectroscopic capability of the RSR on the Large Millimeter Telescope.
The redshifted 492 GHz [C I] line is not detected ($S_{[C~I]}/S_{CO(4-3)} \le 0.45$), but this upper limit is still in line with the measured [C~I] line strengths in other high redshift galaxies \citep[][$S_{[C~I]}/S_{CO(3-2)} \sim 0.3$]{walter11}.

The new CO redshift for COSMOS AzTEC-1 is verified by the detection of redshifted \C2\ line at 335.8 GHz using the SMA.  The bright \C2\ line is detected with $S/N\sim15$, and a higher spectral resolution clearly shows that the line is asymmetric.  The cause of this asymmetry is not known yet, but this explains the slightly lower redshift determined for the \C2\ line.   Although the derived \C2\ line luminosity of $L_{[C\, II]}= 7.8 \times 10^9 L_\odot$ is remarkably high, it is only 0.04 per cent of the total IR luminosity, making COSMOS AzTEC-1 one of the most \C2\ deficient objects known.   We show that AzTEC COSMOS-1 and other high redshift \C2\ sources with a spatially resolved source size extend the tight trend seen between the \C2/FIR ratio as a function of FIR surface density among the IR-bright galaxies by \citet{diaz13} by more than an order of magnitude.  This result lends further support for the explanation that the higher intensity of the IR radiation field and the resulting increased ionization parameter are likely responsible for the ``\C2\ deficiency" seen among luminous infrared starburst galaxies.

Our modeling of the observed spectral energy distribution using a modified black body model, starburst SED models by \citet{efstathiou00}, and the GRASIL SED code \citep{silva98} produces a consistent estimate of the IR luminosity ($L_{IR}=(1.4-1.6) \times 10^{13} L_\odot$ and $L_{FIR}=(0.9-1.2) \times 10^{13} L_\odot$).  The estimated star formation rate from the IR luminosity ($1600-1700\, M_\odot$ yr$^{-1}$) is slightly larger than the model-based $SFR$ (880 and 1320 $M_\odot$ yr$^{-1}$ for the Efstathiou and GRASIL model, respectively).  The model $SFR$ and total stellar mass estimates depend on the adopted star formation history which is intrinsically more uncertain. The best fit GRASIL model constrained by the observed luminosity and the shape of the UV-to-radio SED further suggests of an intense, compact starburst ($r_* \approx 0.1$ kpc) heavily obscured ($A_V>200$ for the molecular clouds and $A_V=3.5$ for the ``cirrus" component) by a massive, compact gas cloud  ($M_{gas}=3.6\pm0.6 \times 10^{11} M_\odot$, $r_{gas}\approx 1$ kpc). 

The total molecular gas mass was derived from the measured CO (4-3) and CO (5-4) lines and 345 GHz continuum using several different methods, specifically addressing the uncertainties associated with each method.  A grid search for non-LTE radiative transfer models that match the observed CO line intensity and line ratio yields acceptable solutions over a wide range of gas temperature and density with a minimum (optically thin) limit of $\sim 10^{10} M_\odot$.  However, plausible models (modest to high optical depth) require a narrow range of gas temperature ($T\approx$ 20-35 K) for densities $n\gtrsim10^{4-5}$ cm$^{-3}$, requiring gas masses of $M_{H2}=(1-7)\times 10^{11} M_\odot$,   Conventional methods of computing molecular gas mass from the observed CO line intensities are subject to very large uncertainties in translating these high $J$ transitions to the intensity of the CO (1-0) line, as well as to the similarly uncertain ``$\alpha_{CO}$" conversion factor.   The empirical calibration that yields a total gas mass of $\sim 2\times 10^{11}M_\odot$ from the measured CO (4--3) line luminosity is on the low end of these different estimates.  The total ISM mass derived from the 345 GHz continuum \citep{scoville14} is near the top of the mass range derived by the other methods: $M_{ISM}=(7.4\pm1.3)\times 10^{11} M_\odot$ for $T_d=25$ K and $(3.6\pm0.7)\times 10^{11} M_\odot$ for $T_d=35$ K.  Future measurements of the CO (1-0) transition should remove the uncertainty associated with the translation of the higher $J$ lines and offer a useful constraint on the CO optical depth.

Our dynamical mass analysis shows that the gas disk in COSMOS AzTEC-1 has to be nearly face-on in order for the derived dynamical mass to be consistent with the minimum possible combined gas and stellar masses.  This offers a natural explanation for the bright, compact stellar light distribution visible in the rest frame UV band \HST\ images, similar to the situation in the local ULIRG Mrk~231.  The same analysis also suggests extremely high opacity ($A_V>200$) for most other viewing angles, as seen in many other high redshift SMGs.  

Among the 15 brightest AzTEC sources identified by the AzTEC/JCMT survey of the COSMOS field and located with a better than 1\arcsec\ positional accuracy using the SMA observations, COSMOS AzTEC-1 is only the second object with a secure spectroscopic redshift, after the $z=5.3$ COSMOS AzTEC-3 \citep{riechers10}.  Advent of the RSR on LMT and other similar broadband spectrometer systems on modern telescopes with a large collecting area (e.g., ALMA) is finally making  accurate determination of redshifts of these distant, optically faint galaxies possible.  In addition to yielding redshifts, these CO spectroscopic surveys can also yield information on total gas masses and dynamical masses along with the excitation conditions of the gas fueling the rapid growths of these young, massive galaxies.  A complete RSR survey of these COSMOS AzTEC sources has started at the LMT, and we should soon be able to gain an unbiased view of the redshift distribution and total molecular gas mass contents of these and other SMGs.

\section*{Acknowledgments}
The authors thank the anonymous referee for the comments and suggestions that improved this manuscript.  The authors also acknowledge the valuable discussions with Lee Armus, Andrew Baker, Daniela Calzetti, Chris Carilli, Giovanni Fazio, Dave Frayer, Andy Harris, Dave Sanders, Nick Scoville, Vernesa Smol{\v c}i{\'c}, Axel {Wei{\ss}}, Al Wootten, and Josh Younger that benefitted this work.  The authors thank R. Blundell for granting us the Director's Discressionary Time (DDT) for the SMA \C2\ observations presented here.
This work would not have been possible without the long-term financial support from the Mexican Science and Technology Funding Agency, CONACYT (Consejo Nacional de Ciencia y Tecnolog\'{i}a) during the construction and early operational phase of the Large Millimeter Telescope Alfonso Serrano, as well as support from the the US National Science Foundation via the University Radio Observatory program, the Instituto Nacional de Astrof\'{i}sica, \'{O}ptica y Electr\'{o}nica (INAOE) and the University of Massachusetts, Amherst (UMass).  The Submillimeter Array is a joint project between the Smithsonian Astrophysical Observatory and the Academia Sinica Institute of Astronomy and Astrophysics and is funded by the Smithsonian Institution and the Academia Sinica.  The UMass LMT group acknowledges support from NSF URO and ATI grants  (AST-0096854, AST-0215916, AST-0540852, and AST-0704966) for the LMT project and the construction of the RSR and AzTEC.  IA, DHH, DSA and MZ«s work is partly supported by CONACyT research grants CB-2009-13326 and CB-2011-167291. DRG is partly supported by CONACyT research grant CB-2011-01-167281. TDS was supported by ALMA-CONICYT grant number 31130005.  RC and HG would like to acknowledge support from a William Bannick Student Travel Grant.  We are grateful to all of the LMT observers from Mexico and UMass who took data for this project. 
This work is based in part on observations made with the Herschel Space Observatory, which is an ESA space observatory with science instruments provided by European-led Principal Investigator consortia and with important participation from NASA, and Planck, which is European Space Agency mission with significant NASA involvement.  This research has made use of the NASA/ IPAC Extragalactic Database (NED) which is operated by the Jet Propulsion Laboratory, California Institute of Technology, under contract with the National Aeronautics and Space Administration.

\appendix

\section{Redshift Determination from an RSR Spectrum \label{sec:appendix}}

\subsection{Template Cross-correlation Analysis}

\begin{figure}
\includegraphics[width=0.99\columnwidth]{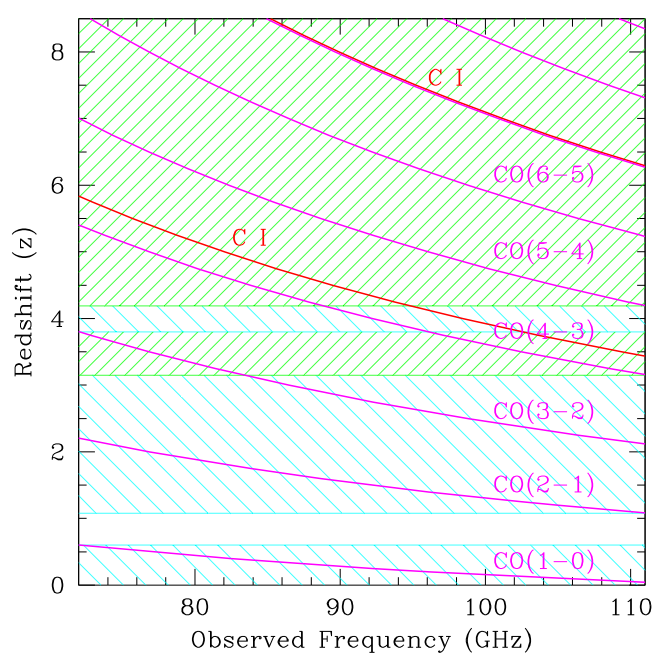}
\caption{Observed frequencies of redshifted CO and [C~I] line transitions falling within the RSR frequency coverage range (73 to 111 GHz).  At least one CO line should appear in the RSR spectrum at all redshifts except for a narrow redshift range of $0.58<z<1.08$.  Two or more CO \& [C~I] lines should appear simultaneously within the RSR spectrum at $z>3.15$.}
\label{fig:COLadder}
\end{figure}

The simultaneous frequency coverage of the RSR between 73 and 111 GHz means at least one CO transition falls within the spectral coverage at all redshifts except for a narrow redshift range between $0.58 <z< 1.08$, and two or more CO or [C~I] transitions fall within the RSR spectral range at $z\ge3.15$ (see Fig.~\ref{fig:COLadder}).  A variety of fainter molecular transitions from less abundant species such as HCN, HCO$^+$, HNC, CS, CN, HC$_3$N, and H$_2$O have also been detected in nearby and distant galaxies \citep[see a review by][and references therein]{carilli13}.  As first introduced by \citet{yun07}, a cross-correlation analysis is a powerful method to derive the redshift information from such a broadband spectrum, even when many of the lines are not individually detected with a good $S/N$ ratio.  A cross-correlation product $\zeta(z)$ can be derived as a function of redshift $z$ from the observed spectrum $S(\nu)$ and the model spectral template $M(\nu,z)$ as
$$\zeta(z)\equiv \int S(\nu) M(\nu,z) W(\nu) d\nu.$$
The Doppler shifted model spectral template  $M(\nu,z)$ is derived as
$$M(\nu,z) = \int_{(\nu-\Delta\nu/2)(1+z)}^{(\nu+\Delta\nu/2)(1+z)} T(\nu')~d\nu'$$
where $T(\nu')$ is the rest frame template spectrum and $\Delta\nu$ is the RSR channel width.  The weight function $W(\nu)$ represents the relative strength of different molecular transitions, and an empirical composite spectrum based on observed relative line strengths for high redshift sources \citep[e.g.,][]{spilker14} is adopted for the analysis presented here.

\begin{figure}
\includegraphics[width=0.99\columnwidth]{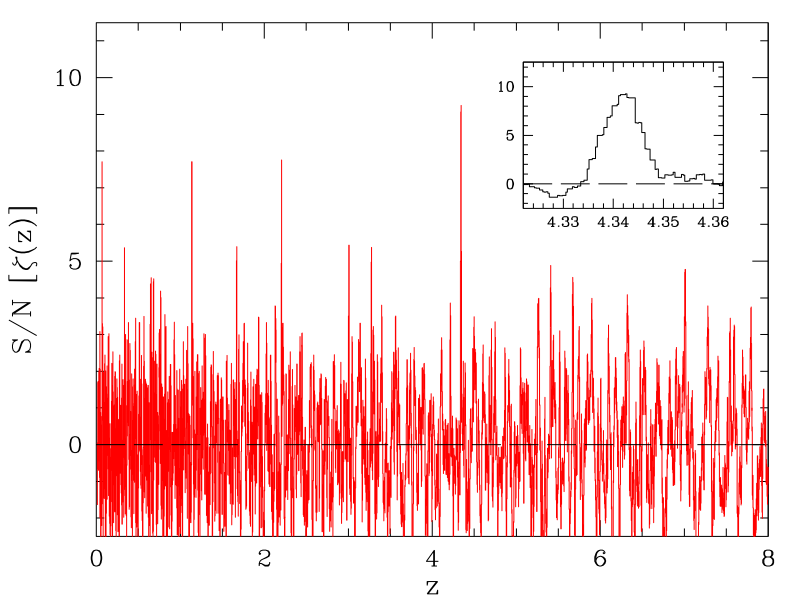}
\caption{Template cross-correlation amplitude of COSMOS AzTEC-1 RSR spectrum in S/N unit.  A zoomed in details of the most significant peak ($S/N=9.0$) at $z=4.342$ is shown in the inset, and it clearly shows that the CO lines are clearly resolved spectrally. }
\label{fig:XCOR}
\end{figure}

The number of spectral lines contributing to the model spectral template $M(\nu,z)$ increases with redshift as the total frequency coverage of the RSR in the rest frame grows as 38$(1+z)$ GHz.  As a result, the noise in the cross-correlation amplitude $\zeta(z)$ increases accordingly with redshift, and interpreting the raw cross-correlation amplitude is not straightforward.  Also, since many of the molecular transitions occurring in the millimetre and sub-millimetre bands are rotational transitions with only slightly different rotation constants, the distribution of line transitions is highly clumped in the spectral domain, further complicating the situation.  Therefore, rather than interpreting the raw cross-correlation amplitude  $\zeta(z)$ for the redshift analysis, we compute a ``$S/N$ ratio" of $\zeta(z)$ for a quantitative analysis of acceptable redshift solutions.  The ``noise" in each redshift bin is estimated by randomly shuffling the input RSR spectrum 10,000 times, and the derived cross-correlation amplitude $\zeta(z)$ is converted to a histogram of $S/N$ ratio as shown in Figure~\ref{fig:XCOR}.  

\subsection{Determination of A Unique Redshift Solution}

The histogram of the template cross-correlation amplitude for the COSMOS AzTEC-1 RSR spectrum in Figure~\ref{fig:XCOR} has the highest peak with $S/N=9$ at $z=4.342$, but other peaks with an apparent $S/N>5$ are also seen.  Since the cross-correlation analysis is sensitive to {\em all} real signal, the two spectral line features detected near 86 GHz and 108 GHz in Figure~\ref{fig:CO} {\em each} produce a series of $S/N$=5-7 peaks that corresponds to different rotational transitions of CO at $z \sim 0$, 1, 2, \& 3, in addition to the strongest peak resulting from the {\em two} CO lines at $z=4.342$.   The presence of two distinct lines for AzTEC-1 in the RSR spectrum (Fig.~\ref{fig:CO}) rules out all single line identifications at $z<3.15$ (see Fig.~\ref{fig:COLadder}), and the $z=4.324$ solution remains as the only plausible interpretation with $S/N>5$.  A small but non-zero possibility of two unrelated CO sources at $z<3.15$ along the same line of sight still remains \citep[e.g.,][]{zavala15}, but we can rule out this scenario by using other tests.

\begin{figure}
\includegraphics[width=0.99\columnwidth]{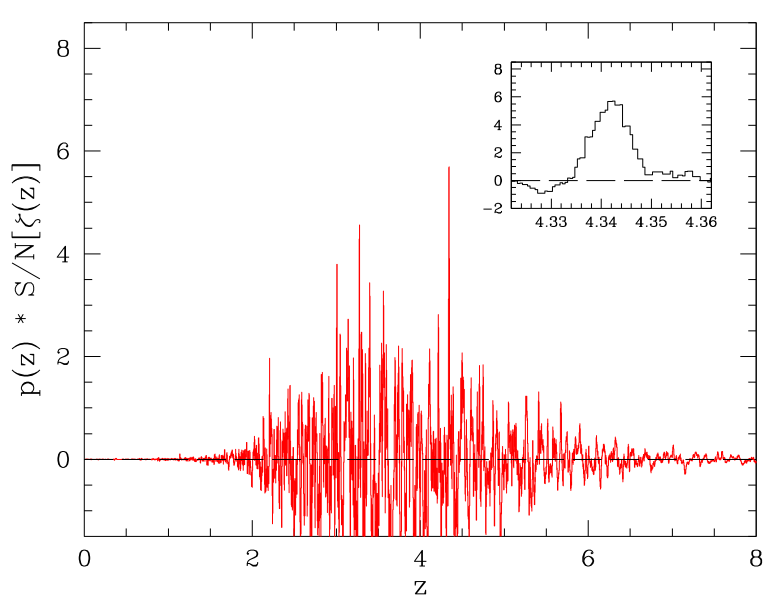}
\caption{A plot of the template cross-correlation amplitude of COSMOS AzTEC-1 RSR spectrum in S/N unit with the redshift constraint from the radio-millimetric spectral index technique \citep{carilli99} utilizing just the AzTEC 1.1mm and the VLA 1.4 GHz photometry.}
\label{fig:XCOR_photz}
\end{figure}

Photometric redshift constraints can be extremely helpful for determining the likely redshift identification \citep{yun07}.  Exploiting the well known radio-IR correlation among star forming galaxies and the strong positive and negative $k$-corrections at the radio and millimetre wavelengths, the radio-millimetric spectral index technique \citep{carilli99} in particular is a simple but remarkably powerful method that requires just two broadband photometry measurements.  As shown in Figure~\ref{fig:XCOR_photz}, the product of the probability distribution for radio-millimetric photometric redshift $p(z)$ and $SNR[\zeta(z)]$ (shown in Fig.~\ref{fig:XCOR}) effectively removes all $z<3$ scenarios and nicely isolates the $z=4.342$ solution.  The effectiveness of the radio-millimetric photometric redshift method is further demonstrated by the fact that a powerful constraint against low redshift solutions can be derived even when only a good radio upper limit is used --  all low redshift solutions can be rejected equally well by treating the $\sim 4 \sigma$ radio photometry point of AzTEC-1 only as an upper limit.

\label{lastpage}

\end{document}